\documentclass[preprint,review,12pt]{elsarticle}

\usepackage[a4paper,margin=1in]{geometry}
\usepackage{setspace}
\makeatletter
\def\ps@pprintTitle{%
\let\@oddhead\@empty
\let\@evenhead\@empty
\def\@oddfoot{}%
\let\@evenfoot\@oddfoot}
\makeatother
\usepackage{graphicx, subcaption}
\usepackage{epstopdf}
\usepackage{color}
\usepackage{multirow}
\usepackage{textcomp,gensymb}
\usepackage{lineno}
\usepackage{xcolor}
\usepackage{physics}
\usepackage{epsfig}
\usepackage{amsmath}
\usepackage{amssymb}
\usepackage{booktabs}
\usepackage{siunitx}
\usepackage{breqn}
\biboptions{square,comma,sort&compress}
\usepackage{hyperref}
\usepackage{subfiles}
\usepackage{credits}
\journal{}

\begin{document}

\begin{frontmatter}
\title{Effect of Magneto-Mechanical Synergism in the Process-Structure Correlation in Fe-C Alloys: A Phase-Field Modeling Approach 
}

\author[a]{Soumya Bandyopadhyay\corref{cor}}
\ead{s.bandyopadhyay@ufl.edu}
\author[b] {Sourav Chatterjee}
\author[c] {Dallas R. Trinkle}
\author[a] {Richard G. Hennig}
\author[a] {Victoria Miller}
\author[d] {Michael S. Kesler}
\author[a]{Michael R. Tonks\corref{cor}}
\ead{michael.tonks@ufl.edu}
\cortext[cor]{Corresponding Authors}

\address[a]{Department of Materials Science and Engineering, University of Florida, Gainesville, Florida-32611, USA} 
\address[b]{Materials Science Division, Lawrence Livermore National Laboratory, Livermore, California 94550, USA} 
\address[c]{Department of Materials Science and Engineering, University of Illinois, Urbana-Champaign, Urbana, Illinois 61801, USA} 
\address[d]{Materials Science and Technology Division, Oak Ridge National Laboratory, Oak Ridge, TN 37831, USA}

\begin{abstract}
Applied magnetic fields can alter phase equilibria and kinetics in steels; however, quantitatively resolving how magnetic, chemical, and elastic driving forces jointly influence the microstructure remains challenging. We develop a quantitative magneto-mechanically coupled phase‑field model for the Fe-C system that couples a CALPHAD‑based chemical free energy with demagnetization-field magnetostatics and microelasticity. The model reproduces single‑ and multi‑particle evolution during the $\alpha\to \gamma$ inverse transformation at 1023 K under external fields up to 20 T, including ellipsoidal morphologies observed experimentally at 8 T. Chemically driven growth is isotropic; a magnetic interaction introduces an anisotropic driving force that elongates $\gamma$ precipitates along the field into ellipsoids, while elastic coherency promotes faceting, yielding elongated cuboidal or ``brick‑like” particles under combined magneto‑elastic coupling. Growth kinetics increase with C content, and decrease with field strength and misfit strain. Multi‑particle simulations reveal dipolar interaction-mediated coalescence for field‑parallel neighbors and ripening for field‑perpendicular neighbors. Incorporating field‑dependent diffusivity from experiment slows kinetics as expected; a first‑principles-motivated anisotropic diffusivity correction is estimated to be small ($<$2\%). These results establish a process-structure link for magnetically assisted heat treatments of Fe-C alloys and provide guidance for microstructure control via chemo‑magneto‑mechanical synergism.
\end{abstract}
\end{frontmatter}
\section{Introduction}
The increasing demand for heat treatment services in the automotive and aerospace industries is driven by the rising production of commercial vehicles and the growing air passenger traffic~\cite{GVR2023}. This growth has prompted the global heat treatment market to develop cost-effective and efficient solutions to enhance materials' performance and durability \cite{GVR2023}.
However, these heat treatments are highly energy-intensive and considered to be the most significant constraint in industrial processes. Despite the industry's maturity, reliance on outdated equipment posses economic, environmental, and operational challenges. For instance, heat treatment of steel consumes trillions of BTUs of energy annually, with a significant loss due to inefficient furnace designs and heat dissipation \cite{andersen2001energy}. Thus, proper optimization of heat treatment and post-processing methods is required to significantly improve energy efficiency.

The recent advancement of superconducting magnets has enabled the generation and application of strong magnetic fields across various domains of materials science and engineering. 
Induction-Coupled Thermomagnetic Processing (TMP) combines high static external magnetic fields with efficient induction to heat treat metals. It reduces process energy consumption of the heat treatment of steels by up to 96\%~\cite{ludtka2013strategic,ludtka2012prototyping,ahmad2012prototyping,ludtka2004situ}. 
Beyond these dramatic energy savings, induction-coupled TMP also accelerates the transformation kinetics {\cite{ludtka2004situ} and improves the tensile strength and hardness of the steels \cite{ludtka2012prototyping}. 


This magnetically-driven phase manipulation finds its origin in the seminal work of Shimotomai \textit{et al.}, who first reported the alignment of paramagnetic fcc austenite ($\gamma$) columns within a ferromagnetic bcc ferrite ($\alpha$) matrix during the $\alpha \to \gamma$ inverse transformation in carbon steels exposed to high magnetic fields \cite{shimotomai2000aligned}.
Furthermore, they mention this type of alignment occurs due to
the dipolar interactions between the nuclei of $\gamma$ phase acting as magnetic holes in the ferromagnetic $\alpha$ phase, resulting in magnetic inhomogeneity within the phases. A concurrent study by Ohtsuka \textit{et al.}~\cite{ohtsuka2000alignment} reported the effect of high magnetic field on $\gamma \to \alpha$ transformation. They observed an accelerated ferrite transformation owing to the increased nucleation rate at a high magnetic field of 10 T. They also report the head to tail alignment of ferrite grains within the austenite matrix along the direction of magnetic field as shown in Fig.~\ref{fig:experiment}(a)-(b). 


\begin{figure}[tbp]
    \centering
        \includegraphics[width=1.0\linewidth]{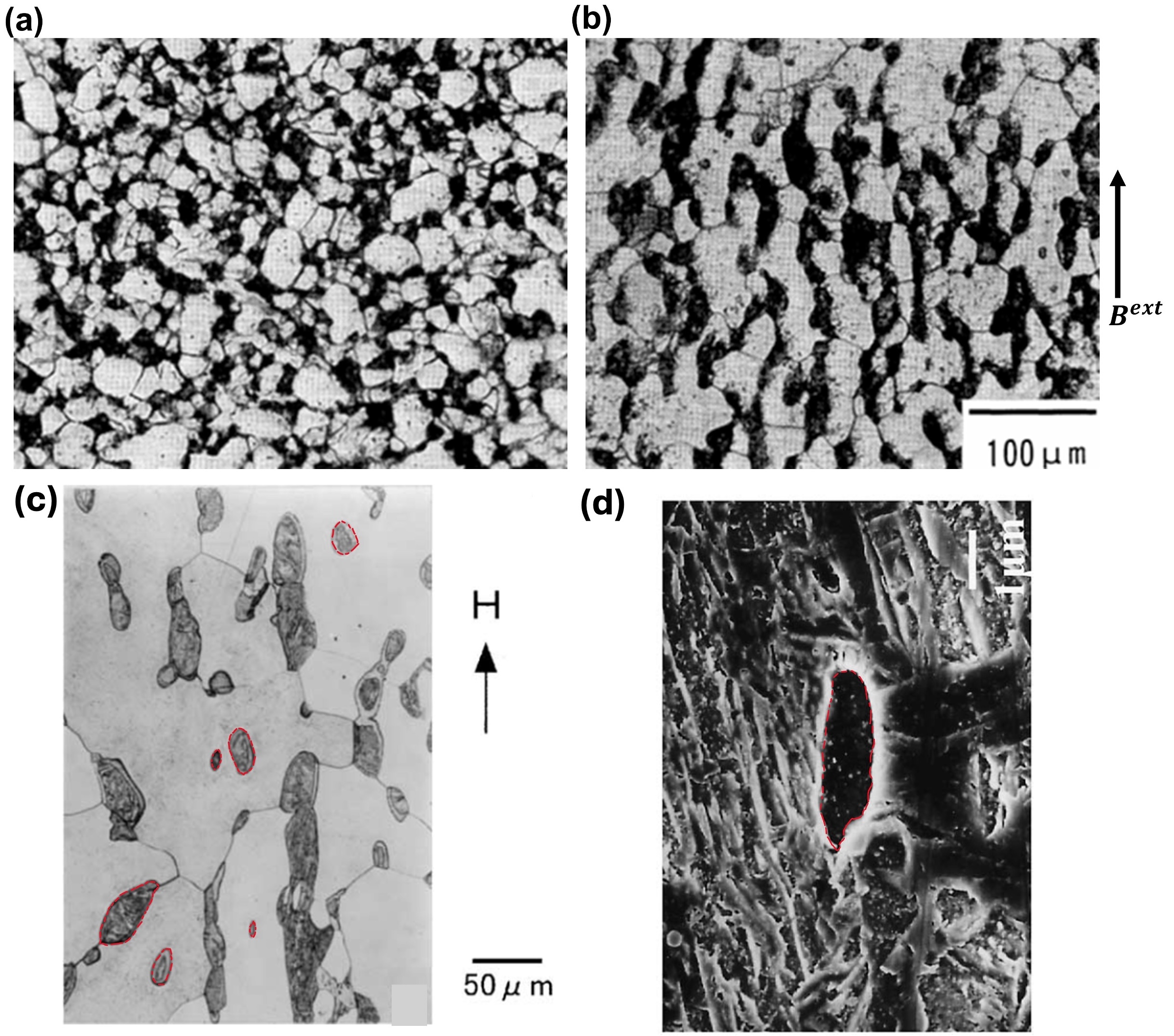}
        \caption{Optical micrographs of Fe–0.4 mass\% C aged at 1023 K, showing microstructure formation:
(a) without applied magnetic field;
(b) with a vertically applied magnetic field.
Black and white regions correspond to the martensite (transformed from $\gamma$) and $\alpha$ phases, respectively. 
The $\gamma$ phase transforms into martensite during quenching. Taken from  Ohtsuka \textit{et al.}
 \cite{ohtsuka2000alignment}. Ellipsoidal grains nucleated in samples with an applied field: (c) microstructures of Fe–0.1\%C alloy subjected to the inverse transformation under magnetic field of 8T. Dark spots are $\gamma$ particles and the
gray matrix is $\alpha$ phase~\cite{maruta2002magnetic}; (d) scanning electron micrographs of steel specimen showing the nucleation of $\alpha$ particles inside a $\gamma$
grain.~\cite{shimotomai2003formation}. Ellipsoidal particles are outlined in red.}
          \label{fig:experiment}
\end{figure}

Additional studies by Shimotomai \textit{et al.} \cite{shimotomai2003formation} on $\gamma \to \alpha$ transformation and by Maruta \textit{et al.} \cite{maruta2002magnetic} on $\alpha \to \gamma$ transformation provide more insights into the impact of the applied field. They found a magnetic field of more than
2T was required to form the aligned morphology \cite{shimotomai2003formation}. 
They also found that the
nucleated grains had an ellipsoidal morphology~\cite{maruta2002magnetic,shimotomai2003formation}, as shown in Fig.~\ref{fig:experiment}(c). Ellipsoidal shapes were also seen when a ferrite particle is nucleated inside an austenite grain (Fig.~\ref{fig:experiment}(d)) \cite{maruta2002magnetic}.
 However, no preferred crystallographic orientations were observed in the aligned grains. Nevertheless, a very recent investigation by 
 Harley et al.~\cite{hurley2025microstructural} reported a lack of alignment of the proeutectoid ferrite morphology at the near eutectoid 
 composition, attributing this to the ferrite’s explicit dependence on the austenite grain boundary network rather than on the applied 
 field. They also mention that slower cooling rates promote morphological 
 elongation~\cite{hurley2025microstructural,wang2007effect,wang2012high}, whereas faster cooling yields less 
 alignment~\cite{zhang2004new}.

Nevertheless, achieving the target microstructure and properties using TMP  still remains challenging, mainly due to the limited understanding of how magnetic fields interact with the chemical and mechanical driving forces in the material to result in the elongated microstructures. Shimotomai \textit{et al.}~\cite{shimotomai2003formation} state that the shape of the particles is not affected by the elastic interaction, if the phases have a nearly similar shear modulus. 
However, the $\gamma$ and $\alpha$ phases
have a small volumetric mismatch \cite{te1998ferrite} which must result in a misfit strain between the two phases that impacts the microstructure.



It is difficult to identify the mechanisms dominating the evolution of the steel microstructure under an applied magnetic field using only experiments, but simulation provides an approach for determining the cause of the observed microstructure. The phase field method  \cite{chen2002phase,steinbach2009phase} has emerged as a powerful tool for modeling the impact of externally applied fields on microstructure evolution\cite{lv2010phase,chafle2019effect,bandyopadhyay2025grain}. 
Koyama and Onodera developed a phase-field model of twin microstructure evolution in Ni$_{2}$MnGa subjected to both stress and magnetic fields~\cite{koyama2003phase}. Later, they extended their model to TMP of $\textrm{Fe–Cr–Co}$ alloys~\cite{koyama2004phase}. Sun \textit{et al.}~\cite{sun2013experimental} developed a model of AlNiCo alloys with an applied magnetic field. Zeng \textit{et al.}~\cite{zeng2022gibbs} developed a phase field model of the formation and elongation of $\gamma$ particles in a $\alpha$ matrix in steels.

In this paper, we attempt to establish the mechanisms causing the elongated microstructures in steels heat treated under an applied magnetic field. We explore the combined impact of the 
chemo-magneto-mechanical synergism on the microstructural evolution using a quantitative phase field modeling approach.
We use a CALPHAD-derived Gibbs free energy for the chemical contribution and incorporate the magnetic and elastic contributions separately in the overall free energy.
We begin in Section \ref{sec:model} by summarizing the phase-field model and its implementation. We then summarize the simulations used to investigate the microstructure evolution in Section \ref{sec:sim_details}. We present our results in Section \ref{sec:results} and discuss their significance in Section \ref{sec:discussion}. Finally, we summarize the key conclusions from our study in Section \ref{sec:conclusions}.

\section{Model Description} \label{sec:model}
We have developed a phase-field model for diffusional phase transformations in the Fe-C system. It includes physically consistent descriptions of individual phases, microelasticity~\cite{khachaturyan2013theory}, and magnetic interactions between different phases~\cite{koyama2003phase,zeng2022gibbs}. Additionally, our model integrates the available CALPHAD database in a thermodynamically consistent manner. In this section, we present the essential modeling components systematically integrated within our diffuse-interface framework. 

In general, there are three types of PF methods for modeling generalized multiphase multicomponent systems: Wheeler-Boettinger-McFadden (WBM)~\cite{wheeler1992phase}, Kim-Kim-Suzuki (KKS)~\cite{kim1999phase}, and grand potential (GP)~\cite{plapp2011unified}. The WBM method, first developed for a two-phase binary solidification \cite{wheeler1992phase} and later extended to multiphase single component systems \cite{steinbach1996phase} and multiphase binary systems~\cite{tiaden1998multiphase}, assumes that the two phases have the same concentration but different volume fraction at a particular point~\cite{wheeler1992phase}. In it, the interfacial width and energy cannot be varied separately. 
In the KKS method~\cite{kim1999phase}, the interface is treated as a mixture of multiple phases, with separate concentration fields with the same chemical potential assigned to each phase. It does allow for the interfacial width and energy to be defined independently, allowing the width to be varied as a modeling parameter.
The GP model was jointly proposed by Plapp~\cite{plapp2011unified} and Choudhury \textit{et al.}~\cite{choudhury2012grand} and is based on the variational principle applied to the grand-potential functional instead of the free energy functional. It is mathematically equivalent to the KKS method but eliminates the need to solve for phase concentrations, reducing the computational cost. However, it generally requires the relationship between concentration and chemical potential be derived from an invertible bulk free energy density~\cite{bognarova2023comparative}. Consequently, despite its higher computational cost, we use the KKS method for our phase field model.


In our model, we represent two phases of the Fe-C system: ferromagnetic ferrite $(\alpha)$ and paramagnetic austenite $(\gamma)$. We use a single non-conserved order parameter field $\eta(\mathbf{r}, t)$, such that $\eta = 1$ in austenite and $\eta = 0$ in ferrite. $0<\eta<1$ across the two-phase interface. We also represent the molar fraction of C using the conserved concentration variable $c(\mathbf{r}, t)$. Here $\mathbf{r} = (x,y,z)$ represents the spatial coordinates and $t$ is time.


\subsection{Thermodynamic Potential} \label{thermopot}
The total free energy of a chemo-magneto-elastic system is expressed as follows~\cite{zeng2022gibbs, koyama2003phase, koyama2004phase, koyama2006phase}:
\begin{equation}
    \mathcal{F} = \int_{V}(f_{bulk} + f_{grad} + f_{elast} + f_{mag}) dV,
  \label{eqn1}
\end{equation}
 where $f_{bulk}$ corresponds to the bulk free energy density of the system and $f_{grad}$ represents the interfacial part. $f_{elast}$ and $f_{mag}$ represent the elastic and magnetic contributions to the system.
%
In our Fe-C model, the bulk and gradient free energy densities for the two
phases are defined as ~\cite{koyama2006phase,zeng2022gibbs,bognarova2023comparative,schwen2017rapid}:} 
\begin{equation}
    f_{bulk} = h(\eta)\frac{f^{\gamma}(c^\gamma)}{V_m} + (1-h(\eta))\frac{f^{\alpha}(c^{\alpha})}{V_m},
\label{eqn4}
\end{equation}
and 
\begin{equation}
    f_{grad} = \omega \eta^2(1-\eta)^2 + \frac{\kappa_{\eta}}{2}\left|\mathbf{\nabla} \eta \right|^2,
    \label{eqn5}
\end{equation}
were $V_m$ is the molar volume and $f^{\gamma}$ and 
$f^{\alpha}$ represent the Gibbs free energy densities of the $\gamma$ and $\alpha$ phases, respectively. 
$h(\eta) = \eta^3(6\eta^2-15\eta+10)$~\cite{schwen2017rapid} 
is an interpolation function that smoothly varies 
between $0$ to $1$ across the phase boundaries, and $\omega$ 
represents the barrier height 
that penalizes wide interfaces.

We define $\omega$ and $\kappa_{\eta}$ is terms of interfacial 
energy $\Gamma$ and interfacial width $l$ as $\omega 
=3K\Gamma/l$ and $\kappa_\eta = \sqrt{6\Gamma l/K}$
\cite{kim1999phase,huang2006phase}, where $K = 2.94$ is a constant that depends on the definition of the interface~\cite{huang2006phase}. The presence of the gradient term 
penalizes sharp interfaces and turns them diffuse.

\begin{figure}
\centering
  \includegraphics[width=0.5\linewidth]{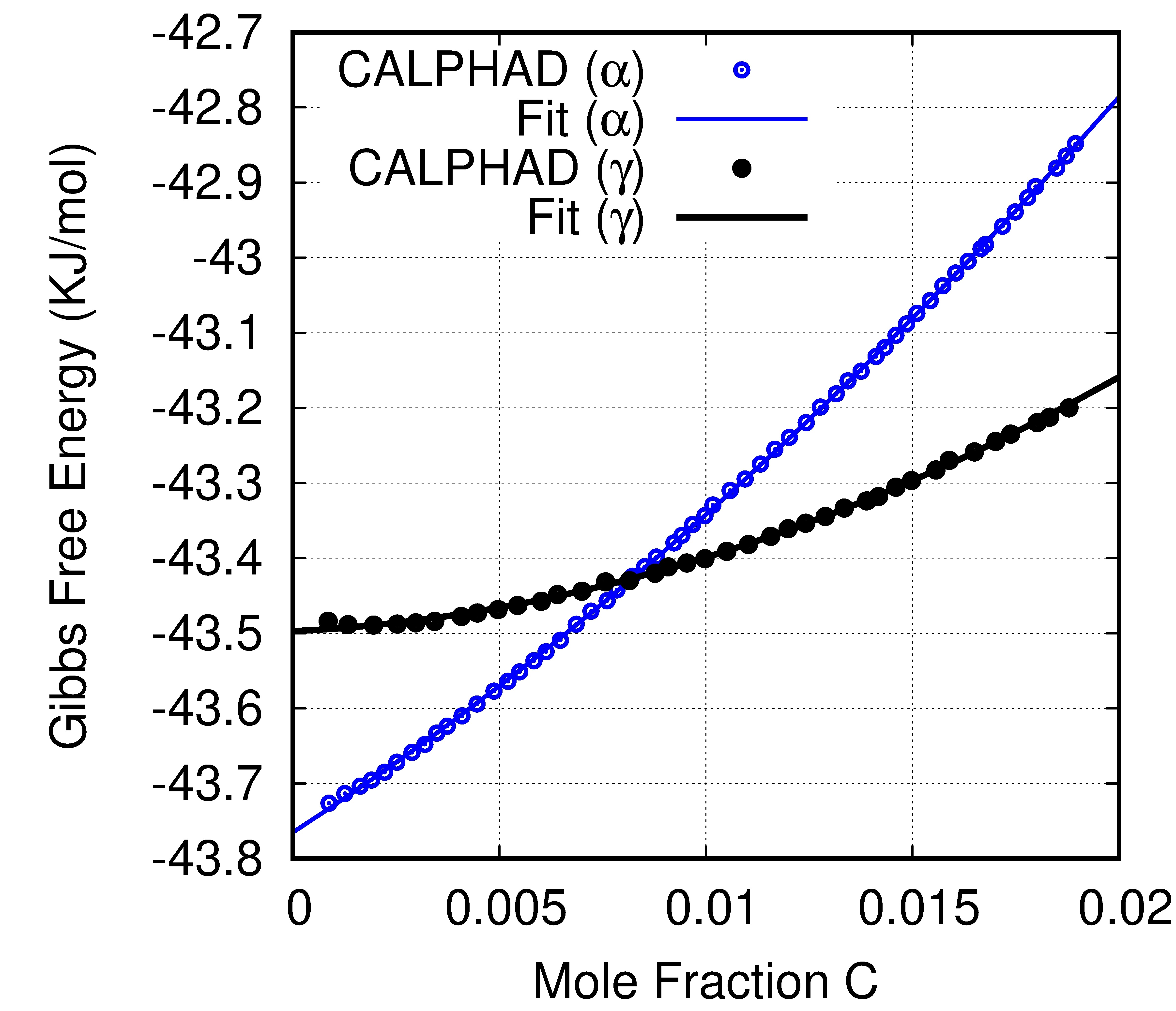}
   \caption{Fitted Gibbs free energies of the $\gamma$ and $\alpha$ phases obtained from CALPHAD databases \cite{naraghi2014thermodynamics,gustafson1985thermodynamic}.}
  \label{fig:gibbs_fit}
\end{figure}

In our present model, we obtain the free energies from CALPHAD databases~\cite{naraghi2014thermodynamics,gustafson1985thermodynamic}
and fit them with the parabolic equations as shown in Fig.~\ref{fig:gibbs_fit}. The forms of the fitted free energies are given as follows:
\begin{subequations}
  \begin{align}
   f^{\gamma} &= \frac{A^{\gamma}}{2}(c^\gamma - c^{\gamma}_{eq})^2,\\
   f^{\alpha} &= \frac{A^{\alpha}}{2}(c^\alpha - c^{\alpha}_{eq})^2,
  \end{align}
  \label{eqn6}
\end{subequations}
where $A^{\gamma}$ and $A^{\alpha}$ are the parabolic coefficients, $c^{\gamma}$ and $c^{\alpha}$ are the compositions of C (as molar fractions) and $c^{\gamma}_{eq}$ and $c^{\alpha}_{eq}$ are the equilibrium compositions in the $\gamma$ and $\alpha$ phases, respectively.
The concentrations $c^{\gamma}$ and $c^{\alpha}$ are local variables that are determined by solving the system of equations defined by the equal chemical potential~\cite{kim1999phase,schwen2017rapid,bognarova2023comparative}:
\begin{equation}
    \frac{\partial f^{\gamma}}{\partial c^{\gamma}} =  \frac{\partial f^{\alpha}}{\partial c^{\alpha}}
    \label{eqn7}
\end{equation}
and 
\begin{equation}
    c = h(\eta) c^\gamma + (1-h(\eta)) c^\alpha.
\end{equation}

\subsection{Magnetic Free Energy}
According to micromagnetic theory \cite{murray20006}, the magnetic free energy contribution consists of five parts~\cite{koyama2003phase,koyama2004phase}: 
\begin{equation}
F_{mag} = \int_{V}(f_{ext} + f_{exch} + f_{an} + f_{str} + f_{d}) dV,
\label{}
\end{equation}
where $f_{ext}$ represents the Zeeman energy density from an external magnetic field, $f_{exch}$ denotes the excess exchange energy density in the magnetic domain wall region, $f_{an}$ accounts for the magnetocrystalline anisotropy energy density, $f_{str}$ the magnetostriction energy density, and $f_d$ represents the magnetostatic or demagnetization energy density. Detailed information on these individual energy contributions can be found in the works of Koyama \textit{et al.}~\cite{koyama2003phase, koyama2004phase, koyama2006phase} and Sun \textit{et al.}~\cite{sun2013experimental}.
The magnetostrictive contribution arising due to coupling between the strain and magnetization is discussed in Section \ref{sec:model-elastic}, along with the elastic contribution.

For the paramagnetic $\gamma$ phase, the total magnetic 
energy is assumed to be zero due to weak magnetization at higher 
temperatures~\cite{zeng2022gibbs}. 
We also neglect the exchange energy and magnetocrystalline anisotropy energy contributions in this study, as 
they do not play a significant role in the alignment of the morphology 
and the overall microstructure~\cite{zeng2022gibbs}.
Moreover, for the $\alpha$ phase, the easy and hard magnetic axes are 
indistinguishable, thus the energy required to deflect the magnetic 
moments from the easy direction to the hard direction is negligible, 
resulting in an 
insignificant anisotropy energy. Therefore, the energy required to rotate magnetic moments between crystallographic directions is negligible, and the intrinsic magnetocrystalline anisotropy can be safely ignored in this context~\cite{zeng2022gibbs}.

However, while the microscopic anisotropy is negligible, the $\gamma/\alpha$ magnetization discontinuity introduces a mesoscopic magnetostatic anisotropy arising from shape effects and interfacial domain alignment. This anisotropy originates from the nonuniform magnetization field that develops near phase boundaries, where surface magnetic charges
produce strong demagnetizing fields and directional driving forces 
that favor morphological elongation or alignment along the field 
direction~\cite{hubert1998magnetic,osborn1945demagnetizing,skomski2008simple}.
Such magnetostatic shape anisotropy is well known to 
dominate microstructural alignment in the
ferromagnetic-paramagnetic systems, even when the intrinsic 
crystalline anisotropy is weak~\cite{skomski2008simple, kittel2018introduction}. 

When a finite-sized ferromagnetic sample is exposed to an external magnetic field, due to the presence of internal magnetic dipole moments, a magnetic flux forms within the sample, generating an additional magnetic field contribution (direction of this field is opposite to the direction of the external magnetic field) known as demagnetization or stray field. It generates an additional energy $F_{d}$ known as the demagnetization energy and is represented as~\cite{cullity2011introduction,zeng2022gibbs}:
\begin{equation}
    f_d = -\frac{1}{2}\mu_0 \mathbf{h}_d\cdot \mathbf{m}(\eta),
\end{equation}
where $\mu_0$ corresponds to the vacuum permeability and $\mathbf{h}_d$ refers to the demagnetization field strength. $\mathbf{m}(\eta)$ is the magnetization of the system and is a function of temperature and phase field order parameter.

Because the present formulation treats the magnetization as a scalar 
field, it captures only the isotropic component of the magnetostatic 
interaction and cannot explicitly resolve these anisotropic 
interfacial field variations. To account for their average energetic 
effect, an effective anisotropy factor $\zeta_{mag}$ was introduced 
and applied solely to the demagnetization term. This correction 
phenomenologically represents the unresolved vectorial and 
interfacial magnetic contributions, ensuring that the magnetic 
driving force attains the correct energetic scale consistent with 
classical demagnetization theory for elongated $\gamma$ inclusions 
embedded in $\alpha$ matrix~\cite{kittel2018introduction, 
osborn1945demagnetizing}.

Accordingly, to take into account of the unresolved anisotropic effects discussed above, the scaled form of this energy term is used as:
\begin{equation}
    f_d^{eff} = -\frac{1}{2}\mu_0 \zeta_{mag}\mathbf{h}_d\cdot \mathbf{m}(\eta),
\end{equation}
Here, $\zeta_{mag}$ represents the effective anisotropy scaling factor applied to the demagnetization term, as discussed above. This modification ensures that the computed magnetic driving force reflects the appropriate energetic scale associated with magnetostatic anisotropy and interfacial field effects. The value of $\zeta_{mag}$ and its derivation are presented in~\ref{apnA}, where it is compared with classical demagnetization energy estimated for elongated $\gamma/\alpha$ inclusions.

 The contribution of the stray energy is obtained by solving the magnetostatic Poisson's equation~\cite{cullity2011introduction,zeng2022gibbs}:
  \begin{align}
    \mathbf{\nabla} \cdot\mathbf{h}_d &= \rho,
    \label{poi}
  \end{align}
where $\rho = -\nabla\cdot \mathbf{m}(\eta)$ denotes the magnetic charge density. According to the magnetic charge method~\cite{zeng2022gibbs}
we can rewrite $\mathbf{h}_d$ as $\mathbf{h}_d = -\nabla \phi$, where $\phi$ is the magnetic scalar potential. 
The contribution from the external magnetic field is incorporated in the model as $f_{ext} = -\mu_0\mathbf{h}_{ext}\cdot \mathbf{m}(\eta)$. Thus, the magnetic driving force is defined as~\cite{zeng2022gibbs}:
\begin{equation}
    \frac{\delta F_{mag}}{\delta \eta } = -\frac{1}{2}\mu_{0}\zeta_{mag}\left(\mathbf{h}_d\cdot \frac{\partial \mathbf{m}(\eta)}{\partial \eta} + \mathbf{m}(\eta)\cdot\frac{\partial\mathbf{h}_d}{\partial \eta}\right),
    \label{magdriv}
\end{equation}
with
\begin{subequations}
  \begin{align}
\frac{\partial \mathbf{m}(\eta)}{\partial \eta} &= \mathbf{m}_\alpha\left(m - \frac{\partial m}{\partial \tau}\tau \right) \textrm{for} \; \tau \leq 1 \\
 & = 0  \;\textrm{for} \;  \tau \ge 1,
\end{align}  
\end{subequations}
where $\mathbf{m}_\alpha$ refers to saturation magnetic flux or in 
some literature saturation magnetization 
\cite{koyama2006phase,zeng2022gibbs}. $\tau = T/T_c$ is the dimensionless temperature normalized by the Curie temperature $T_c$ and $T$ refers to the temperature in which we perform the simulations \cite{koyama2006phase, zeng2022gibbs}.
The form of $m$ or reduced magnetization can be obtained from the analytical description derived from the Weiss-Molecular theory~\cite{barsan2017exact,koyama2006phase, zeng2022gibbs}:
\begin{equation}
\begin{split}
    m(\tau) &= \left(1-2\exp(-2/\tau)\right)\sqrt{1 - \tau}\\
             &\times(1 + 0.60683\tau -0.090480\tau^2 + 5.9531\tau^3 \\
             &- 11.705\tau^4 + 13.950\tau^5 -8.8174\tau^6 + 2.2928\tau^7) \; \textrm{for} \; \tau \leq 1,\\
            & = 0  \; \textrm{for} \; \tau \ge 1.
 \end{split}   
 \label{moment}
\end{equation}
Since in this two phase system only $\alpha$ is 
ferromagnetic, the magnetization $\mathbf{m}(\eta)$ and 
Curie Temperature $T_c$ are considered only with respect to 
$\alpha$ phase by using the interpolation function $h(\eta)$ 
(see Section \ref{thermopot}).

\subsection{Elastic Free Energy}
\label{sec:model-elastic}
The elastic contribution to the free energy is~\cite{khachaturyan2013theory,aagesen2017quantifying}:
\begin{subequations}
  \begin{align}
  F_{elast} &= \int_{V}\frac{1}{2}\boldsymbol{\mathcal{C}}\boldsymbol{\epsilon}_{el}\cdot \boldsymbol{\epsilon}_{el}dV,\\
            &= \int_{V}\frac{1}{2}\boldsymbol{\mathcal{C}}\left(\boldsymbol{\epsilon} - h(\eta)\boldsymbol{\epsilon}^{*}\right)\left(\boldsymbol{\epsilon} -h(\eta)\boldsymbol{\epsilon}^{*}\right)dV,
  \end{align}
  \label{eqn12}
\end{subequations}
where $\boldsymbol{\mathcal{C}}$ is the elasticity tensor, $\boldsymbol{\epsilon}_{el}$ is the elastic strain tensor, and $\boldsymbol{\epsilon}$ and $\boldsymbol{\epsilon}^{*}$ are the total strain and the stress-free or eigen strain tensors, respectively. The elasticity tensor $\boldsymbol{\mathcal{C}}$ varies depending on the order parameter $\eta$ according to $\boldsymbol{\mathcal{C}} = h(\eta)\boldsymbol{\mathcal{C}}^{\gamma} + (1-h(\eta))\boldsymbol{\mathcal{C}}^{\alpha}$, where $\boldsymbol{\mathcal{C}}^{\gamma}$ and $\boldsymbol{\mathcal{C}}^{\alpha}$ denote the elasticity tensor for the $\gamma$ and $\alpha$ phases, respectively. We assume linear elasticity such that $\boldsymbol{\epsilon} = 1/2 (\nabla \mathbf{u} + \nabla \mathbf{u}^T)$, where $\mathbf{u}$ is the displacement vector.

The eigenstrain tensor consists of two parts~\cite{khachaturyan2013theory}:
$\boldsymbol{\epsilon}^{*} = \boldsymbol{\epsilon}^{mis} + \boldsymbol{\epsilon}^{mstr}$. Here,  $\boldsymbol{\epsilon}^{mis}$ corresponds to the epitaxial or misfit strain arising due to the lattice parameter mismatch between the phases and  $\boldsymbol{\epsilon}^{mstr}$ represents the magnetostrictive strain that refers to the spontaneous lattice deformation
coupled to the local magnetization.
Physically, the lattice parameters of each phase and the lattice misfit are functions of composition~\cite{aagesen2017quantifying}. Using a constant misfit strain 
tensor dependent solely on the order parameter fails to capture this compositional dependence~\cite{aagesen2017quantifying}. Moreover, introducing compositional dependence may introduce an additional composition-dependent strain gradient, further complicating the equation-solving process. To mitigate this complexity, we have adopted an order-parameter-dependent misfit strain tensor $\boldsymbol{\epsilon}^{mis} = \epsilon_{ls} \mathbf{I}$, where $\epsilon_{ls}$ is the scalar lattice strain and $\mathbf{I}$ is the identity tensor. The form of magnetostrictive strain is given as~\cite{koyama2003phase,zhang2005phase} 
\begin{equation}
 \begin{aligned}
   \boldsymbol{\epsilon}^{mstr} =\begin{cases}  \frac{3}{2}\lambda_{100}\left(m_i m_j - \frac{1}{3}\right), & (i = j)\\
                                     \frac{3}{2}\lambda_{111} m_i m_j, & (i \neq j),
    \end{cases}                                 
  \end{aligned}
  \label{eqnmagstr}
\end{equation}
where $\lambda_{100}$ and $\lambda_{111}$ are the magnetostrictive coefficients and $\mathbf{m}$ is the magnetization vector. Since the magnetostrictive 
term in Eq.~\ref{eqnmagstr} is much weaker ($\lambda_{100} \approx \num{20e-06}$~\cite{cullity2011introduction}) than the 
coherency strain, it does not exert substantial qualitative influence on the overall morphology. Therefore, the magnetostrictive 
contributions are neglected in our simulations.

Finally, to obtain the elastic energy, we have to solve for the local stress tensor $\boldsymbol{\sigma} = \boldsymbol{\mathcal{C}} \boldsymbol{\epsilon}$ throughout the domain. This is done by solving the quasistatic mechanical equilibrium equation:
\begin{equation}
    \nabla \cdot \boldsymbol{\sigma} = 0.
    \label{mech}
\end{equation}

\subsection{Kinetic Equations}
Once we obtain all the energetic contributions, the spatio-temporal evolution of the carbon composition $c$ is obtained by solving the Cahn-Hilliard equation as~\cite{cahn1958free}:
\begin{equation}
    \frac{\partial c}{\partial t} = \mathbf{\nabla} \cdot M_c \mathbf{\nabla}\left( \frac{\delta F}{\delta c}\right),
    \label{ch}
\end{equation}
where $M_c$ refers to the compositional mobility which varies with the local 
phase and is constructed by interpolation via the function $h(\eta)$ as: $M_c = M_c^{\alpha}(1-
h(\eta)) + M_c^{\gamma}h(\eta)$. The single‐phase mobilities $M_c^{i}$ (with $i$ = $\alpha$ , 
$\gamma$) are related to their respective diffusivities $D^{i}$ by
\begin{equation}
    M_c^{i} = D^i/\frac{\partial^2 f^i}{\partial {c^{i}}^2}.
\end{equation} 
Finally, the diffusivities in both the 
$\gamma$ and $\alpha$ phases are assumed to follow an Arrhenius‐type temperature dependence: 
$D^{i} = D^i_{0}\exp\left(\frac{Q_{i}}{RT}\right)$,
where $D^i_0$ is the prefactor and $Q_i$ is the 
activation energy for diffusion in phase $i$.

The non-conserved order parameter $\eta$ evolves by the Allen-Cahn equation as follows~\cite{allen1979microscopic}:
\begin{equation}
    \frac{\partial \eta}{\partial t} = - L \left( \frac{\delta F}{\delta \eta}\right),
    \label{ac}
\end{equation}
where $L$ denotes the mobility of the interface which can be obtained from 
sharp interface analysis~\cite{kim1999phase} and 
bulk diffusional mobility $M_c$ through $M_c/l^2$, where $l$ denotes the interfacial width 
\cite{karma1998quantitative, echebarria2004quantitative}. In our simulations $L$ is 
set large enough for the kinetics to be diffusion-controlled~\cite{aagesen2017quantifying}.  

\section{Simulation details} \label{sec:sim_details}
\subsection{Model Implementation in MOOSE}
In this section, we detail the procedure employed to solve the system of equations defined in the previous section. 
We solve the system of partial differential equations mentioned in the previous section using the finite element method (FEM) and implicit time integration with the Mulitphysics Object-Oriented Simulation Environment (MOOSE), an open source FEM framework \cite{permann2020moose}. We determine the compositional evolution using Eq.~\eqref{ch} and phase evolution using Eq.~\eqref{ac}. The magnetic scalar potential ($\phi$) as well as the demagnetization field ($\mathbf{h}_d$) is obtained using Eq.~\eqref{poi}. Furthermore, we use Eq.~\eqref{mech} to solve for the displacement vector $\mathbf{u}$ and the strain $\boldsymbol{\epsilon}$.

The resultant system of non-linear equations is solved using preconditioned Jacobian Free Newton Krylov~\cite{aagesen2017quantifying, gaston2009moose,schwen2017rapid} with a second-order backward Euler differentiation scheme~\cite{gaston2009moose}. We use automatic non-linear residual scaling and set the relative and absolute convergence criterion to be $10^{-8}$ and $10^{-10}$, respectively. We employ adaptive time stepping with an initial time step size of $0.01$ with roughly $15$ non-linear iterations per time step. We make use of additive Schwarz domain decomposition that partitions the non-linear system of equations into sub-blocks and then uses a LU preconditioner for each sub-block.

We use first-order Lagrange QUAD4 elements for the simulations. For the elastic solver, we use the Global Strain Formalism \cite{biswas2020development} that captures macroscopic volume changes, shear deformations, and other uniform strains while preserving periodic strains across all boundaries. Periodic boundary conditions are used for all other variables.

\subsection{Simulations setup}
In our microstructural evolution simulations, we choose an interfacial width of $l = 5\delta$, where $\delta = \SI{6}
{\nano\meter}$ is the smallest grid spacing. We employ dynamic mesh adaptivity to concentrate resolution 
where it matters most to reduce the computational cost.  At each time step, a local error is computed for every mesh element, flagging regions with large gradients (typically 
around interfaces) for refinement and regions with no gradients for coarsening. 

The interfacial energy was assumed to be $\SI{0.5}{\joule\per\meter^2}$~\cite{mecozzi2008role}. All parameters corresponding to the materials properties and simulations are listed in Table~\ref{table1}. References are provided where applicable.

\begin{table}[btph]
\begin{center}
\caption{Numerical parameters used for the phase field simulations with references provided where applicable.}
\label{table1}
\begin{tabular}{||c|c||} 
 \hline
 Parameters & Values  \\ [0.5ex] 
 \hline\hline
  System dimension  & $\SI{1}{\micro\meter} \times \SI{1}{\micro\meter}$   \\ 
  \hline
  $\Gamma$  & $\SI{0.5}{\joule\per\meter^2}$~\cite{mecozzi2008role}   \\ 
  \hline
   Magnetic field (\textbf{B}) & $1$T, $10$T,  $8$T $20$T  \\
   \hline
    Effective anisotropic factor ($\zeta_{mag}$) & $10$~\ref{apnA} \\
   \hline
    misfit strain $\epsilon_{ls}$ & $0.3\%,-0.3\%,-1.0\%, -2.0\%, -3.0\%$  \\
     \hline
    $C_{11}^{\gamma}$, $C_{12}^{\gamma}$, $C_{44}^{\gamma}$ & $\SI{206}{\giga\pascal}$, $\SI{134}{\giga\pascal}, $\SI{113.5}{\giga\pascal}~\cite{kim2016evaluation} \\
    \hline
    $C_{11}^{\alpha}$, $C_{12}^{\alpha}$, $C_{44}^{\alpha}$ & $\SI{221.9}{\giga\pascal}$, $\SI{143.6}{\giga\pascal}, $\SI{113.8}{\giga\pascal}~\cite{kim2016evaluation} \\
    \hline
    Curie Temperature ($\alpha$ phase) $T_c^{\alpha}/(K)$ & \SI{1042}{\kelvin}~\cite{koyama2006phase,zeng2022gibbs} \\
\hline
 Aging Temperature $T (K)$  & \SI{1023}{\kelvin}\\
\hline
Saturation magnetic flux ($\alpha$ phase) $M_{\alpha}$(T) & 2.22~\cite{koyama2006phase,zeng2022gibbs} \\
\hline
Alloy composition & Fe- x mass\%C (x = 0.1-0.4)\\
\hline
$D_0^{\alpha}$, $Q_0^{\alpha}$ & \num{1.31e-06}$m^2/s$, \SI{81.4} {\kilo\joule\per\mol}~\cite{lan2004modeling}\\
\hline
$D_0^{\gamma}$, $Q_0^{\gamma}$ & \num{1.7e-05}$m^2/s$, \SI{143.32}{\kilo\joule\per\mol}~\cite{lan2004modeling}\\
\hline
$L$ & \num{1.0e-02} $m^3/Js$ \\
\hline
$A^{\alpha}$, $A^{\gamma}$ &  \num{1.1278e06}, \num{1.2377e06} J/mol\\
\hline
\end{tabular}
\end{center}
\end{table}

\section{Results} \label{sec:results}

We apply our phase field model to investigate the mechanisms causing the elongated particles observed in experiments as shown in  Fig.~\ref{fig:experiment}. We explore the impact of the elastic and magnetic energies. 
We begin by validating our study using the conditions analogous to those studied by Maruta \textit{et al.}~\cite{maruta2002magnetic}, where austenite grains nucleate within a ferrite matrix during the inverse transformation at 1023 K. 
Then we perform parametric analysis 
with simulations of single particle growth followed by simulations with multiple particles. 
Our simulations attempt to comprehend the transformation pathway, with a  
particular focus on the growth of austenite within the ferrite matrix under the influence of magneto-elastic coupling. We end by including the impact of an applied magnetic field on the diffusivity in the model.

\subsection{Validation with experiments} \label{sec:results_single_validation}
First we validate our model by reproducing studies similar to the 
classic 8 T experiment of Maruta \textit{et al.}~\cite{maruta2002magnetic}. 
As mentioned earlier,
during the diffusional transformation the $\gamma$ and $\alpha$ phases have slightly different 
atomic volumes, producing a small volumetric misfit ($\approx1\%$ \cite{te1998ferrite}) but 
essentially no shear. This yields an isotropic compressive linear eigenstrain of $0.3\%$. Shimotomai 
\textit{et al.}~\cite{shimotomai2003formation} noted that the strain energy associated with an 
ellipsoidal precipitate depends weakly on shape because the shear moduli of the phases are 
nearly identical, but even when the nearly similar elasticity tensors, lattice mismatch can influence
the precipitate morphology. 

Thus, we consider a scenario in which, during the inverse transformation, a single austenite precipitate nucleates within the ferrite matrix. To perform the simulations, we employ a domain size of $\SI{1}{\micro\meter} \times \SI{1}{\micro\meter}$. We begin with a configuration consisting of a single circular precipitate of austenite ($\gamma$) embedded in a ferrite ($\alpha$) matrix. The initial precipitate radius is $\SI{35}{\nano\meter}$, as shown in Fig.~\ref{fig:evol}(a). The initial molar fraction of carbon in the austenite and
ferrite phases are 0.027 and 0.00074, respectively. 
 We use $\epsilon_{ls}=-0.003$ and an applied field of 8 T. The nucleus is allowed to grow until it reaches steady state.

\begin{figure}[tbp]
    \centering
    \begin{subfigure}[t]{1.0\textwidth}
        \centering
        \includegraphics[width=1.0\linewidth]{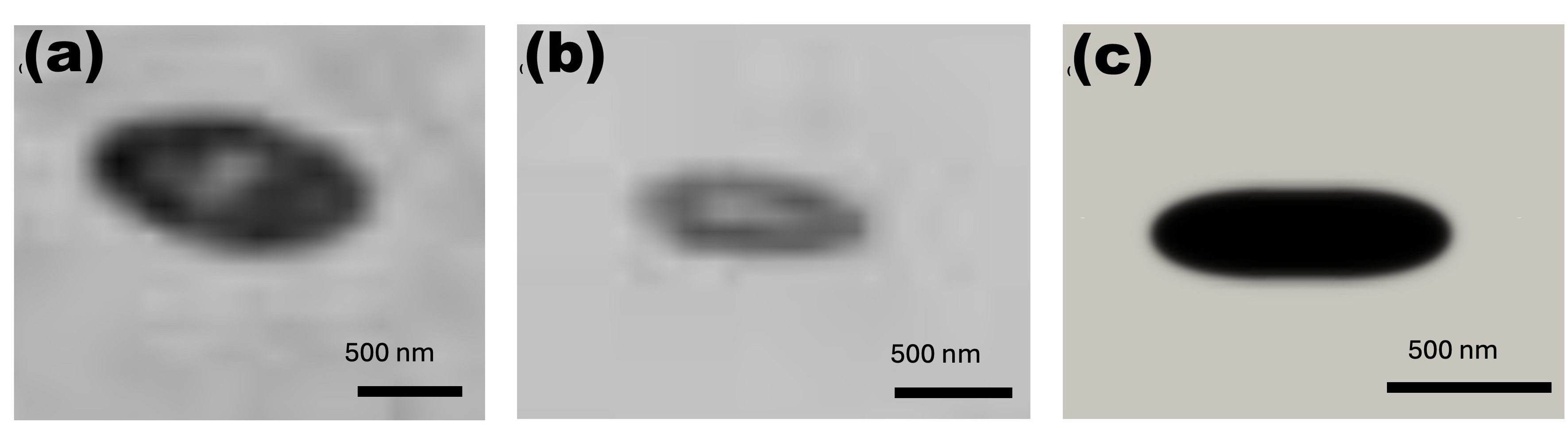}
    \end{subfigure}%
    \caption{(a)-(b) Representative sections of the microstructures (shown in Fig.~\ref{fig:experiment}(c)) obtained from the experimental report by Maruta \textit{et al.}~\cite{maruta2002magnetic} for Fe-0.1 \%C under 8T magnetic field. (c) Simulated morphology for the same system with similar experimental conditions and misfit strain of $0.3\%$. The simulated morphology displays an elliptical shape under these condition, which agrees well with the experimental observation.}
    \label{fig:expcompa}
\end{figure}
Figures~\ref{fig:expcompa} (a)-(b) represent the morphology of $\gamma$ particles (dark regions) 
embedded in the $\alpha$ phase (light region) for the Fe-0.1\%C system under 8 T. 
Figure~\ref{fig:expcompa}(c) displays the simulated 
morphology for the same system, where we observe an elliptical microstructure similar to what 
was seen in the experiment. This similarity demonstrate that our magnetoelastic coupled model reproduces a well-characterized experimental case.

\subsection{Single particle system} \label{sec:results_single}
After completing the validation experiments, we delve deeper into the magnetic and 
elastic effects on the precipitate morphology. Accordingly, we conduct 
parametric studies to explicitly isolate and assess the impact of the individual energy component. We use the same domain size, initial circular nucleus, and initial concentrations as from the validation simulation.  In this case, we consider
a far-field supersaturation value $c_\infty = 0.0182$ for the Fe-0.4 wt\%C system. We systematically investigate the influence of different energy contributions on the evolution of the microstructure. Specifically, we consider: (a) chemical contributions (bulk and interfacial), (b) chemical and magnetic contributions, (c) chemical and elastic contributions, and (d) combined chemo-magneto-mechanical contributions.

 \begin{figure}[tbp]
\centering
 \includegraphics[width=
 1.0\linewidth]{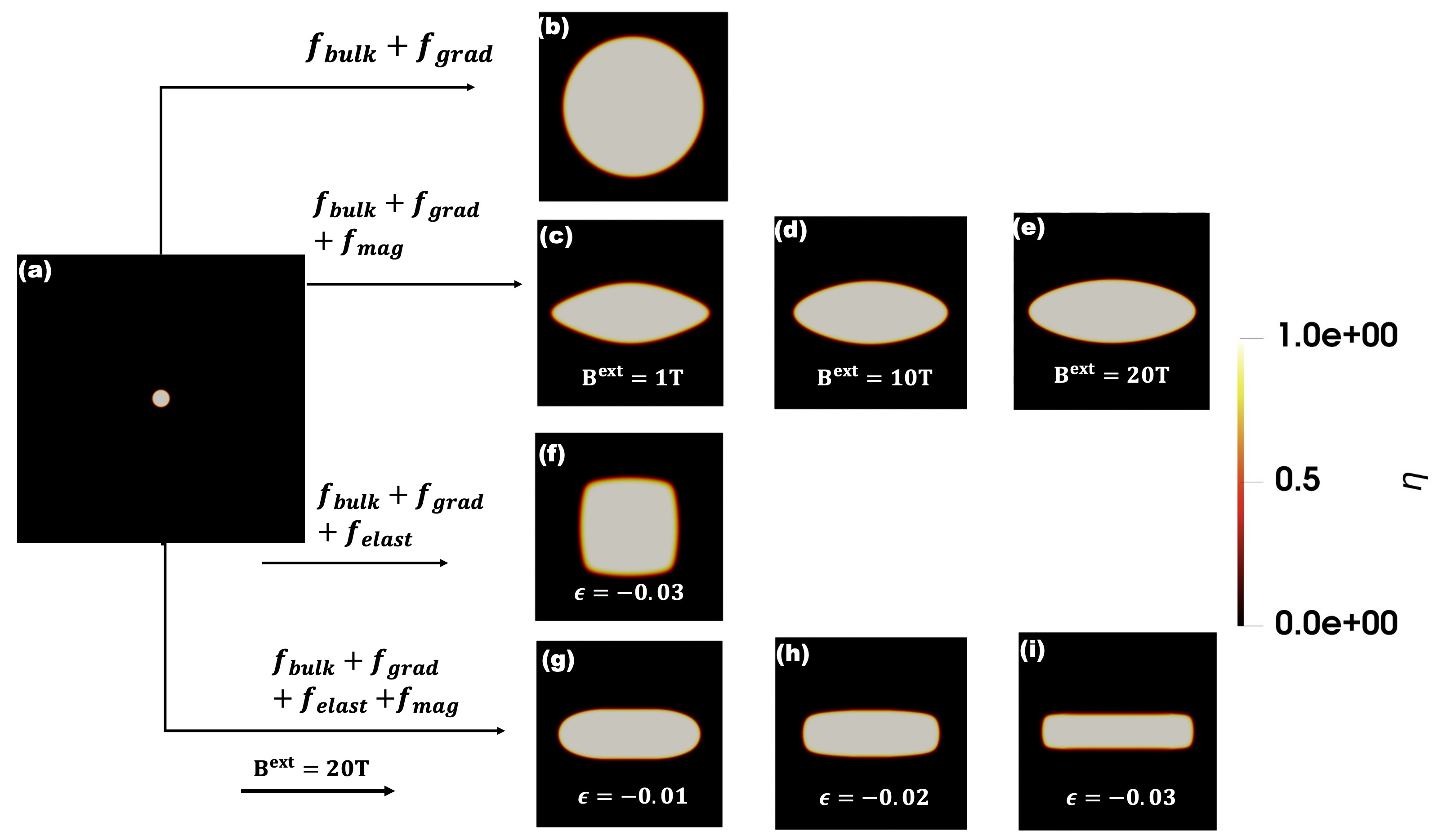}
 \caption{Microstructural evolution of an austenite precipitate within a ferrite matrix in a Fe-0.4 wt\%C system:
(a) initial configuration;
(b) shape resulting from chemical effects alone, showing an isotropic profile;
(c–e) evolution under 1 T, 10 T, and 20 T magnetic fields, transitioning from a lens shape at 1 T to a perfectly elliptical morphology at higher fields;
(f) profile arising from combined chemical and elastic interactions, yielding a cuboidal precipitate;
(g–i) combined magneto-elastic effects at 20 T, evolving from a faceted ellipse at low misfit strain to a brick-like morphology at higher misfit strain.}
\label{fig:evol}
\end{figure}%

In the absence of magnetic and elastic effects, the precipitate grows isotropically, as illustrated in Fig.~\ref{fig:evol}(b), driven solely by bulk and interfacial energies, yielding a circular morphology.
Figures~\ref{fig:evol}(c)–(e) show the effect of magnetic field on the microstructural evolution in Fe-0.4\% C system. At $1~\textrm{T}$, the precipitate begins to elongate, adopting a lens-shaped form. As the field strength increases to $10~\textrm{T}$ and $20~\textrm{T}$, it further transforms into a well-defined ellipse.

Figure~\ref{fig:evol}(f) displays the final morphology under an applied epitaxial compressive misfit strain of $3\%$, where the precipitate adopts a cuboidal shape. This morphology is frequently observed in superalloys~\cite{kommel2015phase}. 

When the effects of magnetic and elastic fields are combined with the chemical contributions, the resulting microstructures exhibit intriguing morphological features. Figures~\ref{fig:evol}(g)–(i) show the final morphologies with a high magnetic field of $20~\textrm{T}$ and compressive misfit strains of $1\%$, $2\%$, and $3\%$. 
Figure~\ref{fig:evol}(g) illustrates the microstructure under a $-1\%$ misfit strain. Here, we observe an elliptical morphology oriented along the horizontal axis due to the magnetic field, but the vertical edges are faceted due to the elastic field. As the misfit strain increases to $-2\%$ and $-3\%$, shown in Figs.~\ref{fig:evol}(h) and (i), the microstructure retains its elongated form, but develops more pronounced faceting in both horizontal and vertical directions, resulting in an elongated cuboid or brick-like appearance. 

\begin{figure*}[tbph]
    \centering
    \begin{subfigure}[t]{0.92\textwidth}
        \centering
        \includegraphics[width=1.05\linewidth]{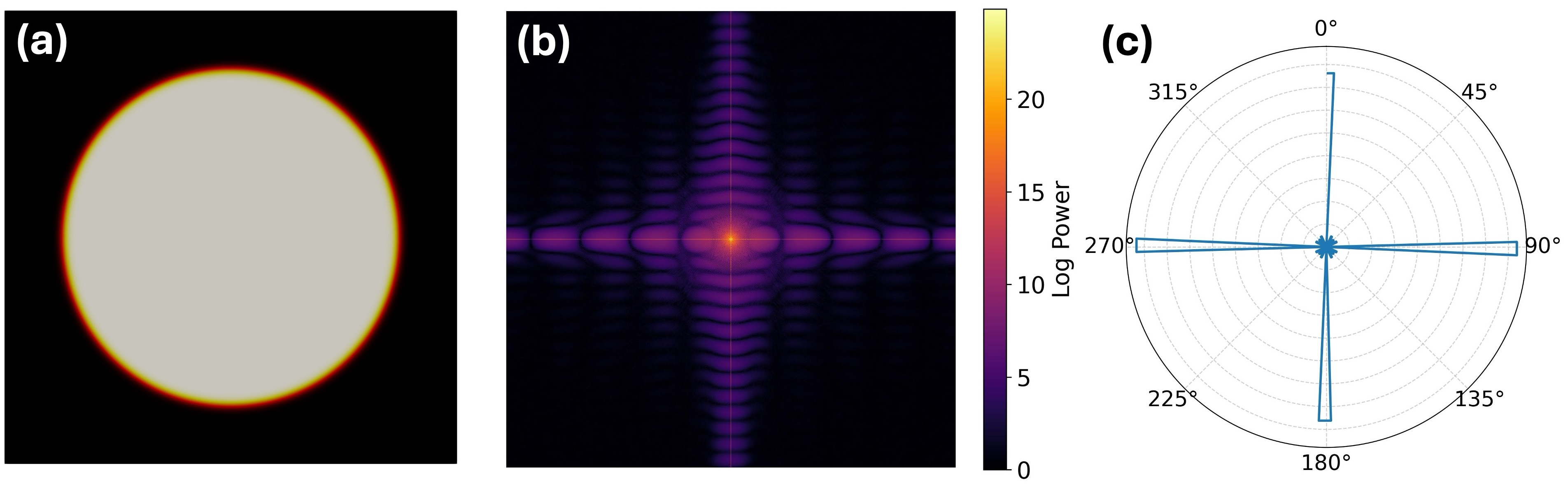}
    \end{subfigure}%

    \begin{subfigure}[t]{0.92\textwidth}
        \centering
        \includegraphics[width=1.05\linewidth]{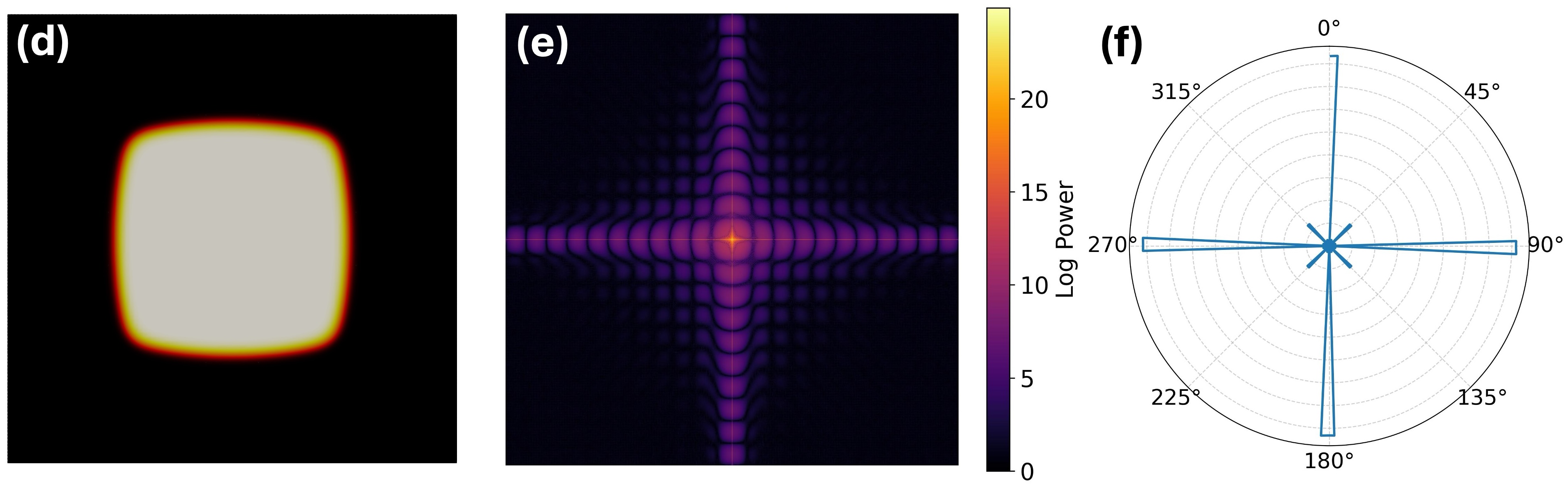}
    \end{subfigure}
   
   \begin{subfigure}[t]{0.92\textwidth}
      \centering
        \includegraphics[width=1.05\linewidth]{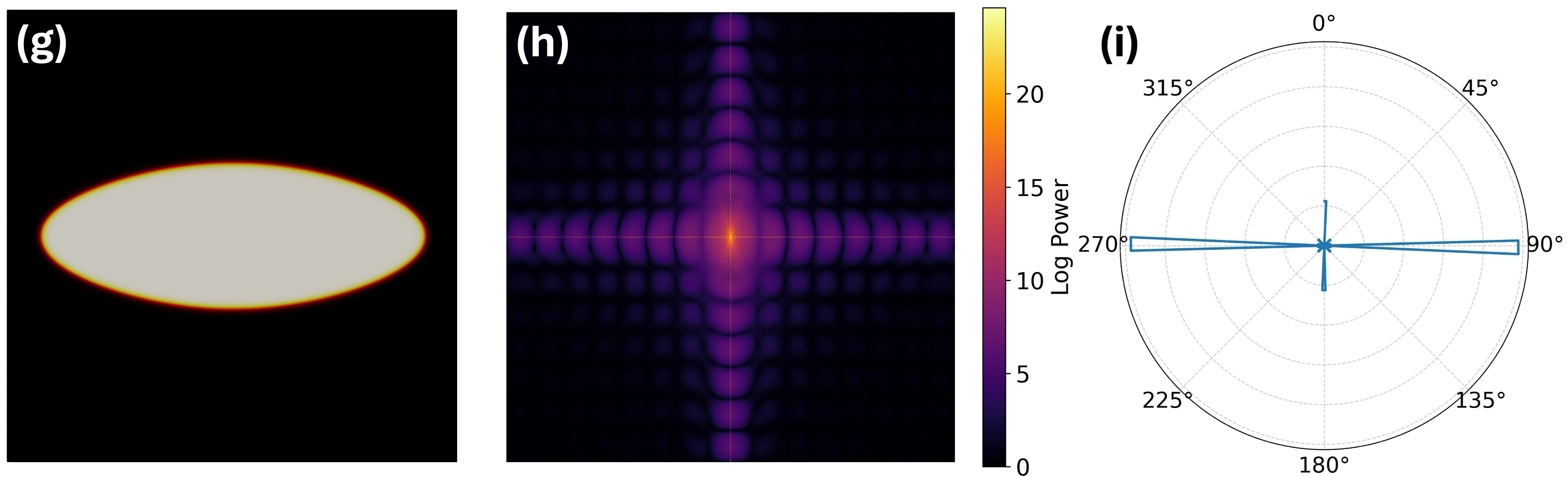}
    \end{subfigure}
   
     \begin{subfigure}[t]{0.92\textwidth}
      \centering
        \includegraphics[width=1.05\linewidth]{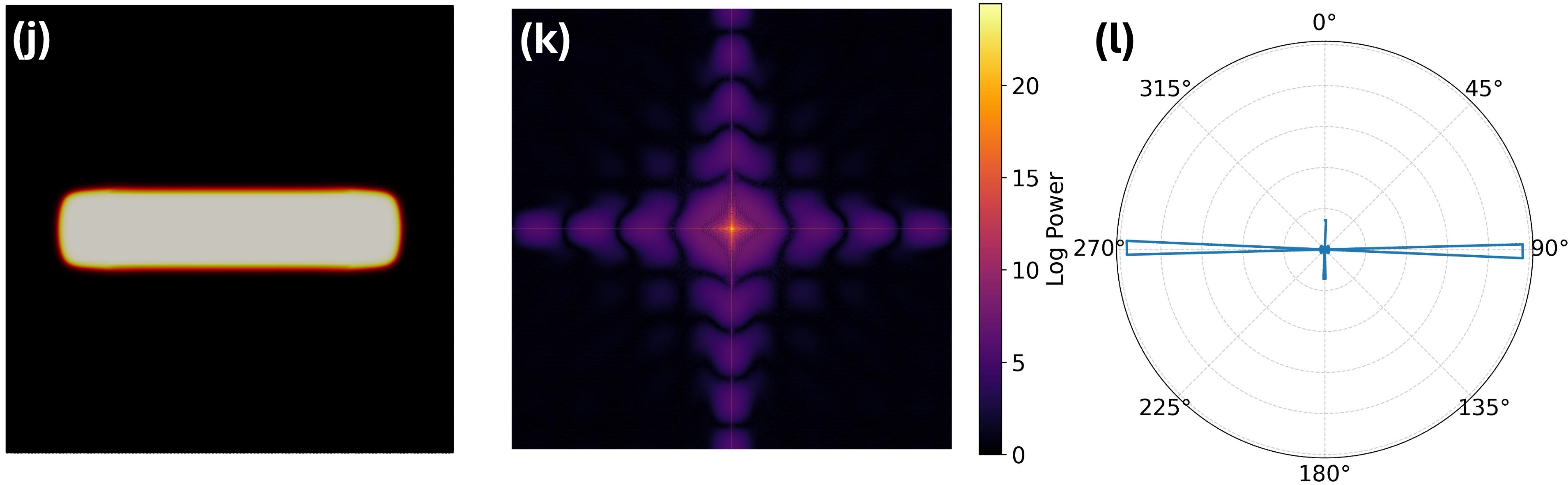}
    \end{subfigure}
    \caption{Power spectrum analysis (b,e,h,k) and angular power distribution (c,f,i,l) for  austenite precipitates (a,d,g,j) in a Fe-0.4 wt\%C system for (a-c) chemical effects along, (d-e) elastic‐only interactions with $\epsilon_{ls} = -0.03$, (g-i) magnetic interaction with $B^{ext} = 20$ T, and (j-i) under combined magnetic and elastic interactions with $\epsilon_{ls} = -0.03$ and $B^{ext} = 20$ T.}
    \label{fig:fft}
\end{figure*}
A more quantitative approach of comparing the precipitate morphologies is with Fourier power spectrum analysis using Fast Fourier Transform (FFT). This is a well-established microstructure characterization technique,
widely used to uncover the hidden symmetries, dominant spatial frequencies, and 
preferred orientations \cite{ tanaka1986application,loren2002determination}. It is analogous to small angle X-ray scattering
(SAXS) data \cite{tanaka1986application,zhu2018microstructure}. Figures~\ref{fig:fft}(a)-(c) show the FFT analysis for the system with only the chemical contribution, revealing an isotropic spectrum with no preferred orientation. Introducing elastic interactions (Fig.~\ref{fig:fft}(d)) produces a similar isotropic FFT response (Figs.~\ref{fig:fft}(e)-(f)). We observe a similar scenario, despite the underlying cubic anisotropy that produces a cuboidal precipitate because the applied misfit strain is itself isotropic or dilatational. In contrast, when we apply the magnetic field along the horizontal direction, the particle elongates along the direction of the applied field (Fig.~\ref{fig:fft}(g)). The corresponding FFT analysis also shows a similar response with increased power at $90\degree$ and $270\degree$ as displayed in Figs.~\ref{fig:fft}(h)-(i). 
When both magnetic and elastic effects are present (Fig.~\ref{fig:fft}(j)) the precipitate morphology differs, but the FFT signature remains the same as shown in Figs.~\ref{fig:fft}(k)-(l).

\begin{figure*}[tbp]
    \centering
    \begin{subfigure}[t]{0.49\textwidth}
        \centering
        \includegraphics[width=1.0\linewidth]{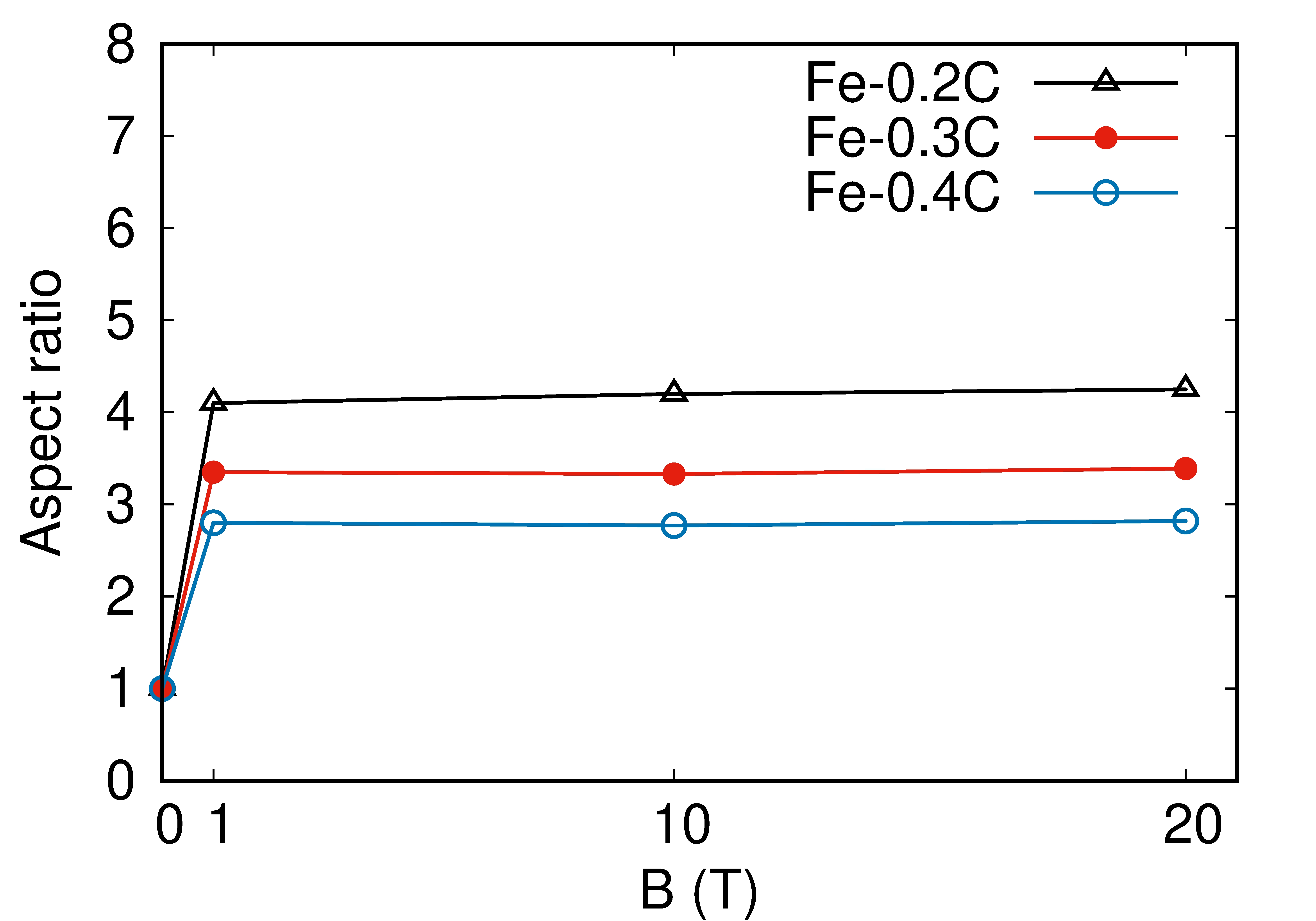}
        \caption{}
          \label{fig:regmag}
    \end{subfigure}%
    \begin{subfigure}[t]{0.49\textwidth}
        \centering
        \includegraphics[width=1.0\linewidth]{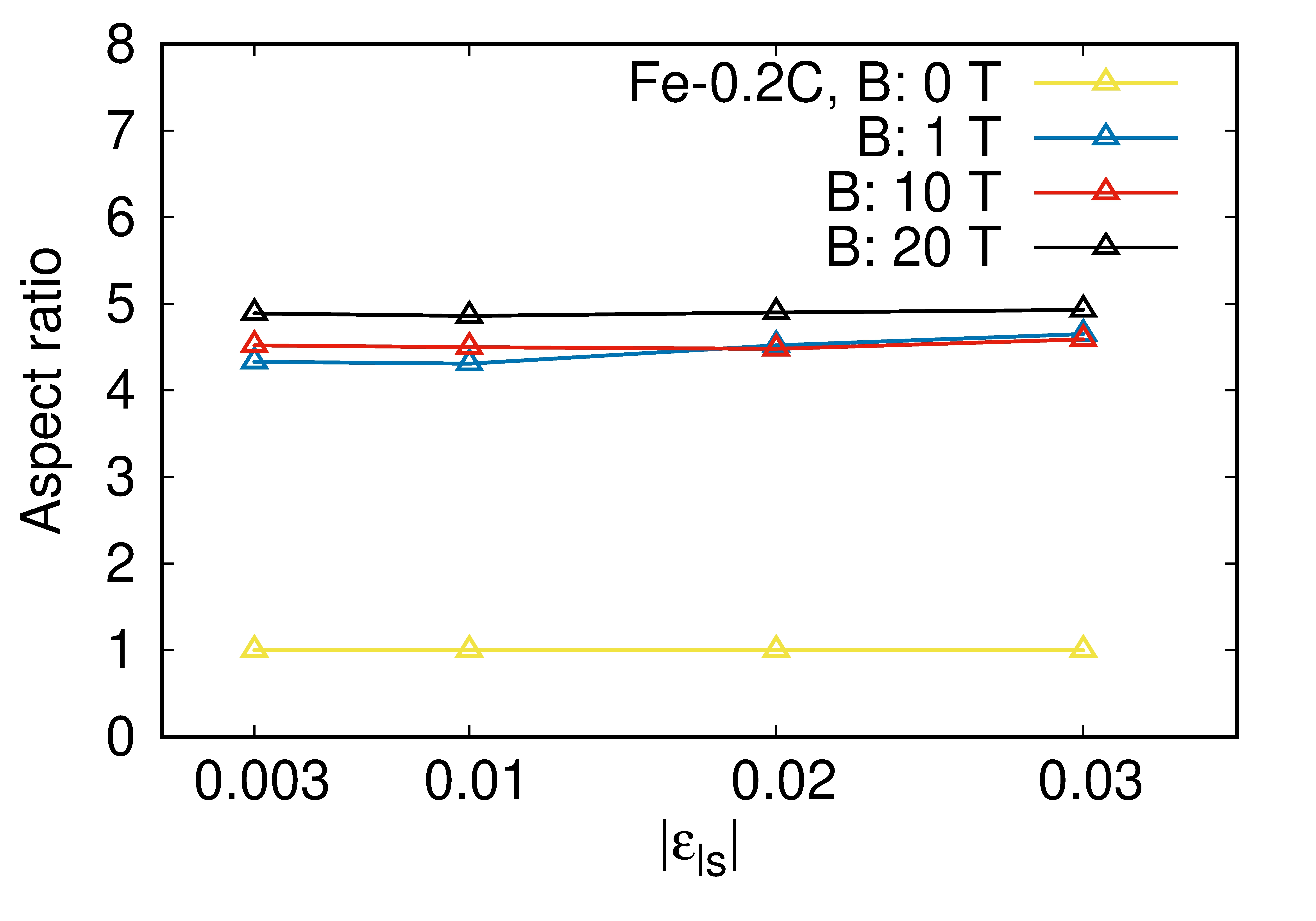}
        \caption{}
        \label{fig:regstr}
    \end{subfigure}
 \begin{subfigure}[t]{0.49\textwidth}
        \centering
        \includegraphics[width=1.0\linewidth]{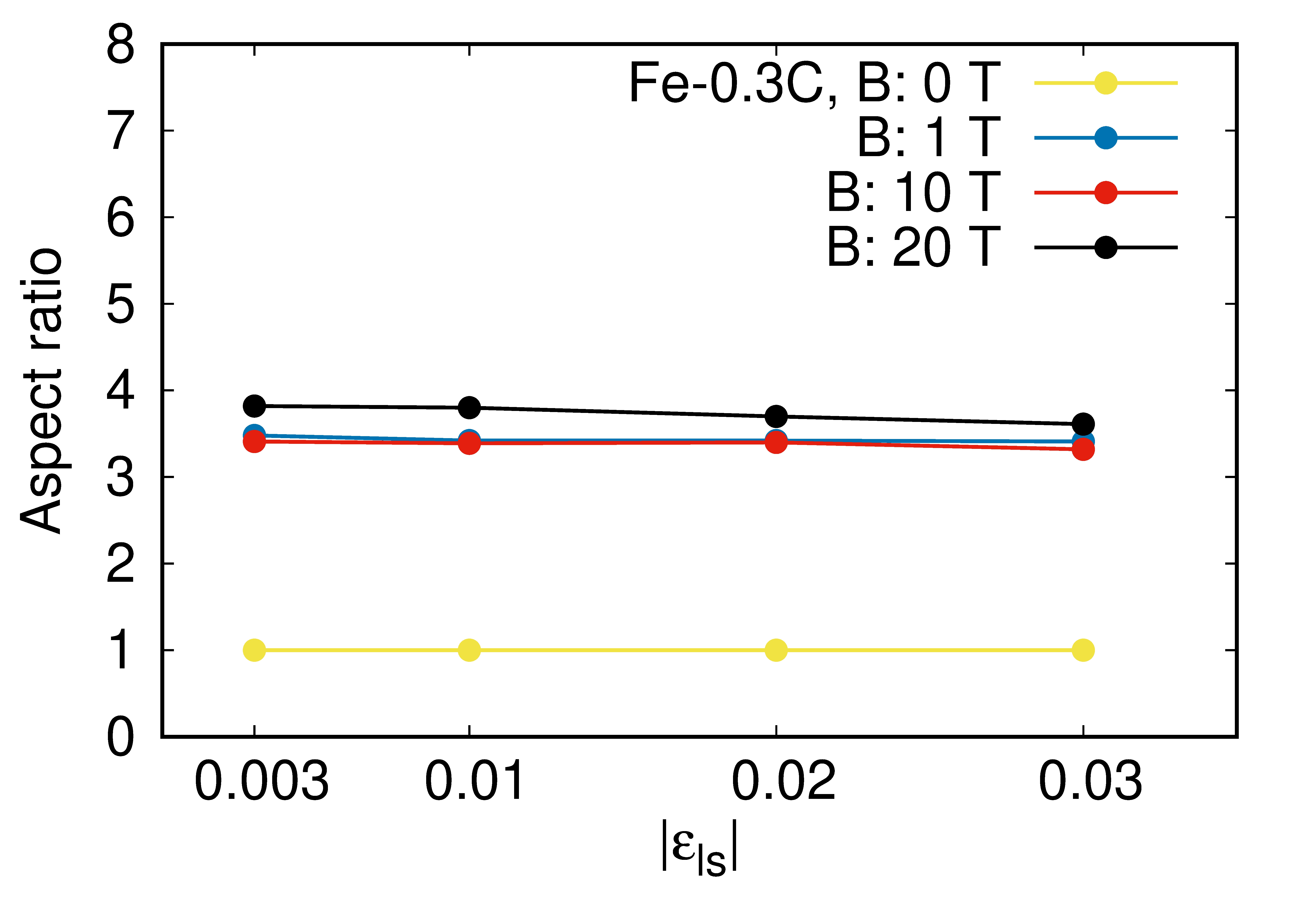}
        \caption{}
        \label{fig:regstr0.3}
    \end{subfigure}
    \begin{subfigure}[t]{0.49\textwidth}
        \centering
        \includegraphics[width=1.0\linewidth]{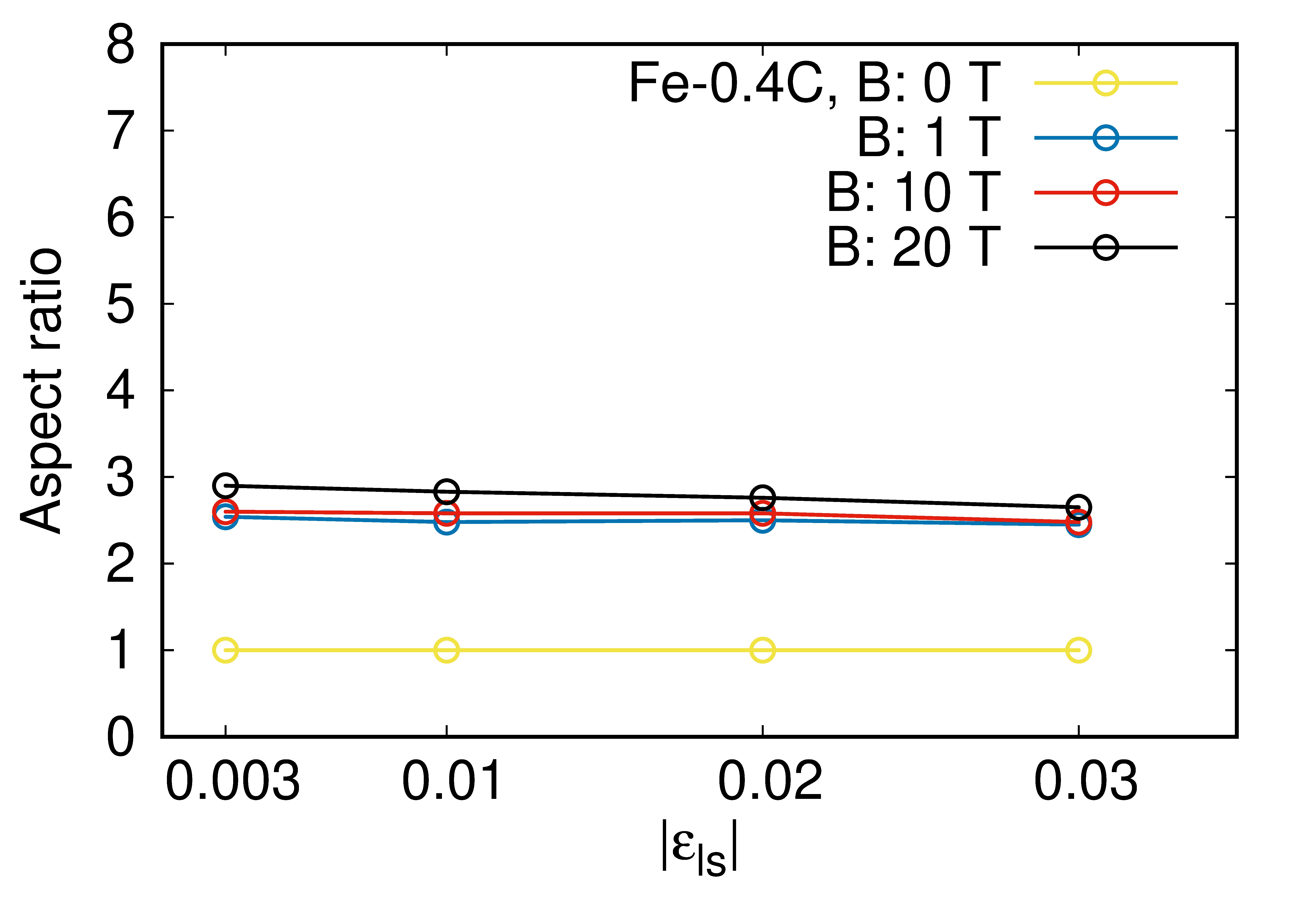}
        \caption{}
        \label{fig:regstr0.4}
    \end{subfigure} 

\caption{Change in precipitate aspect ratio as a function of (a) applied magnetic field for different carbon concentrations,  
(b) - (d) misfit strain for Fe-0.2 wt\%C, Fe-0.3 wt\%C, and Fe-0.4 wt\%C, respectively, for different magnetic fields.}
     \label{fig:aspect}
\end{figure*}
 
 We have also carried out an additional study focusing on the aspect ratio ($AR$) of the growing precipitate for various eigenstrains, alloy compositions, and applied magnetic field.
 Figure~\ref{fig:regmag} shows how 
the aspect ratio varies with magnetic field and 
composition. We observe a sharp increase from $AR=1.0$ with 0 T to values ranging from around 2.5 to 4.2, depending on the composition. Additional increases in applied field from 1 T to 20 T result in very small linear increases in aspect ratio. The magnitude of the impact of the applied field on the aspect ratio decreases with increasing carbon concentration. 

Figures~\ref{fig:regstr}-~\ref{fig:regstr0.4} display the variation of the aspect ratio as a function of misfit strain $\epsilon_{ls}$ for different compositions and magnetic fields. In these data, we observe a trend similar to that discussed previously.
While, the aspect ratios decrease with increasing compositions, they marginally increase with increasing magnetic field strength. Furthermore, 
in the presence of coupled magneto-elastic effects the aspect ratios exhibit a sharp jump from the isotropic case ($AR=1$ with 0 T) to higher values ($AR\approx 2.5 - 5.0$ with $\geq0$ T) as discussed before. 

There is little impact of the misfit strain on the aspect ratio, consistent with the FFT analysis from Fig.~\ref{fig:fft}. For Fe-0.2\%C (see Fig.~\ref{fig:regstr}), there is a slight increase in aspect ratio with misfit strain, for Fe-0.3\%C (see Fig.~\ref{fig:regstr0.3}) there is little effect, and for Fe-0.4\%C (see Fig.~\ref{fig:regstr0.4}) there is a slight decrease. 
Although the overall phase transformation is intricately governed by the thermodynamics and kinetics, in all cases the aspect ratios lie between 2.5 and 5. These values are in excellent agreement with the values measured and calculated (between 2.0 and 4.9) by Shimotomai \textit{et al.}~\cite{shimotomai2003formation}.

\begin{figure*}[tbph]
    \centering
    \begin{subfigure}[t]{0.5\textwidth}
        \centering
        \includegraphics[width=1.0\linewidth]{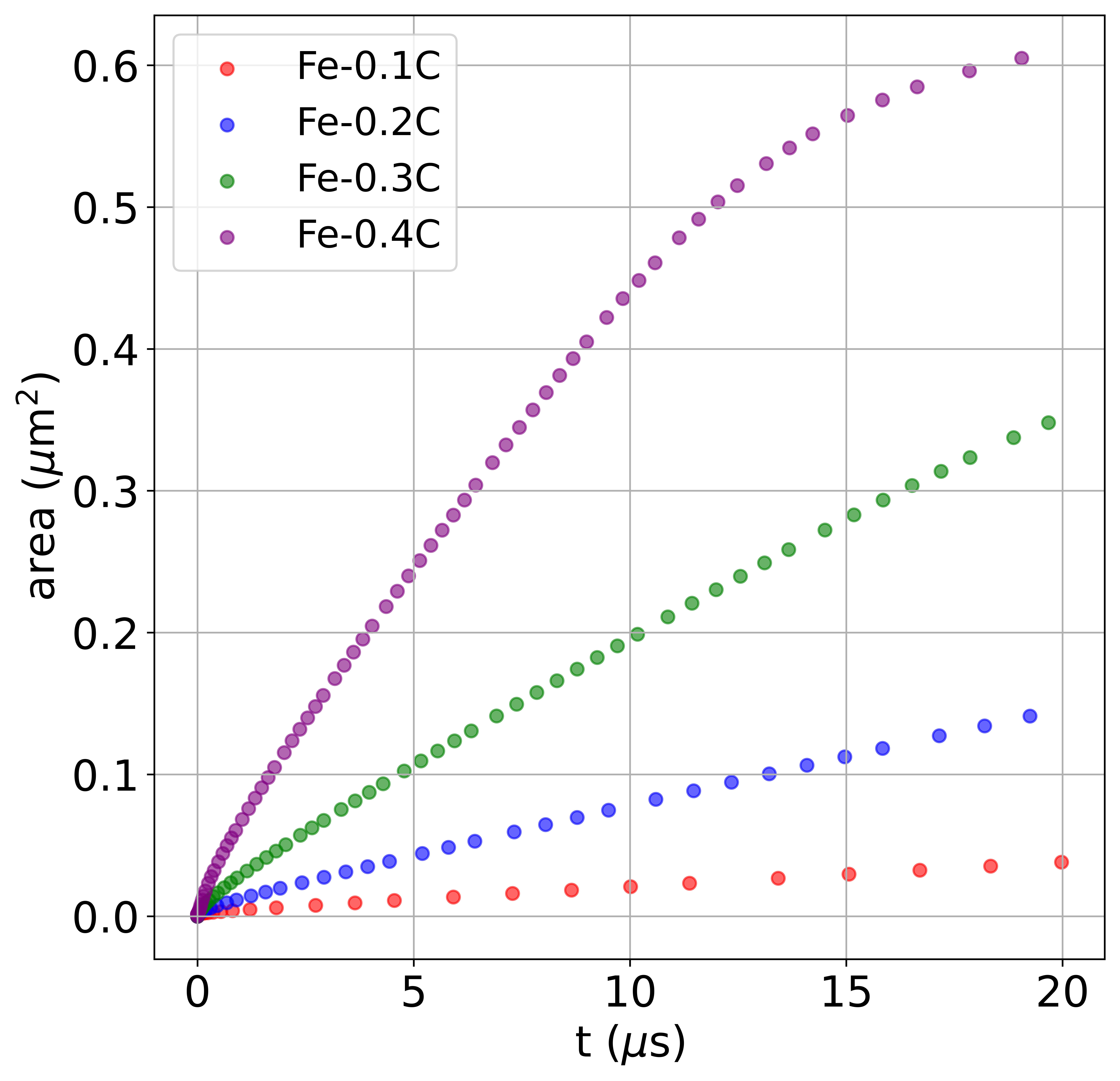}
        \caption{}
    \end{subfigure}%
    \begin{subfigure}[t]{0.49\textwidth}
        \centering
        \includegraphics[width=1.0\linewidth]{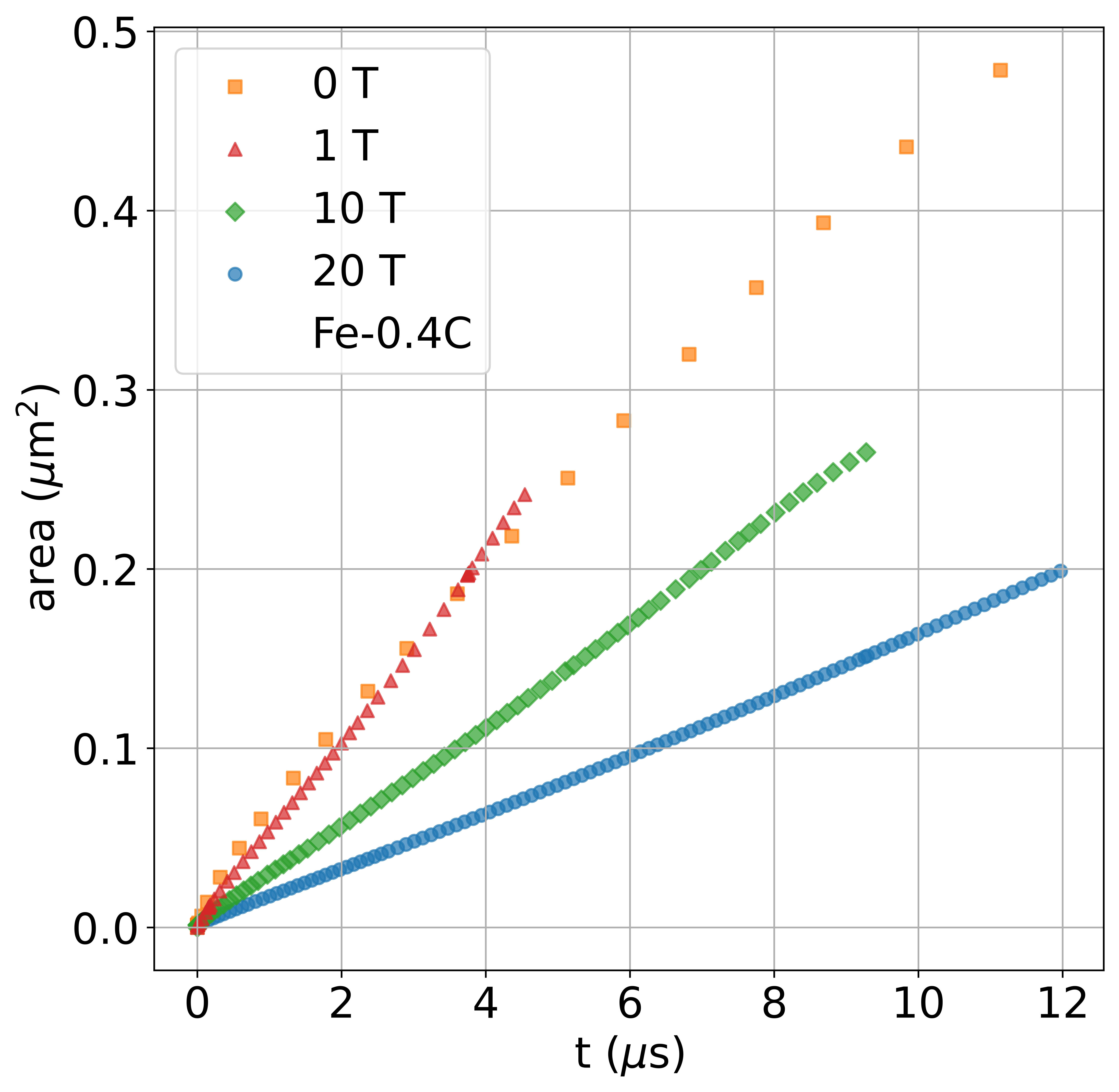}
        \caption{}
    \end{subfigure}
   \begin{subfigure}[t]{0.49\textwidth}
      \centering
        \includegraphics[width=1.0\linewidth]{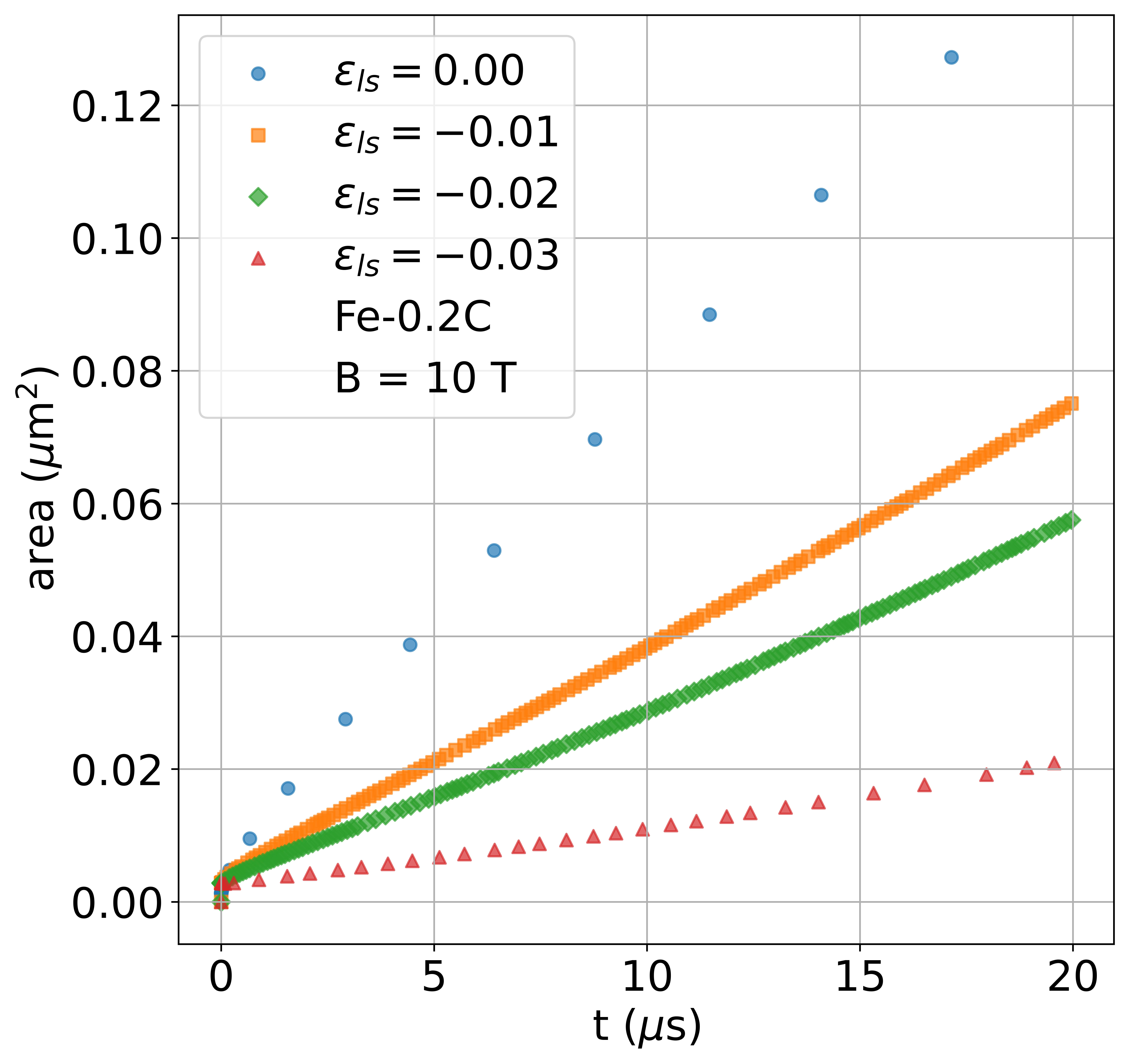}
        \caption{}
    \end{subfigure}
     \begin{subfigure}[t]{0.49\textwidth}
      \centering
        \includegraphics[width=1.0\linewidth]{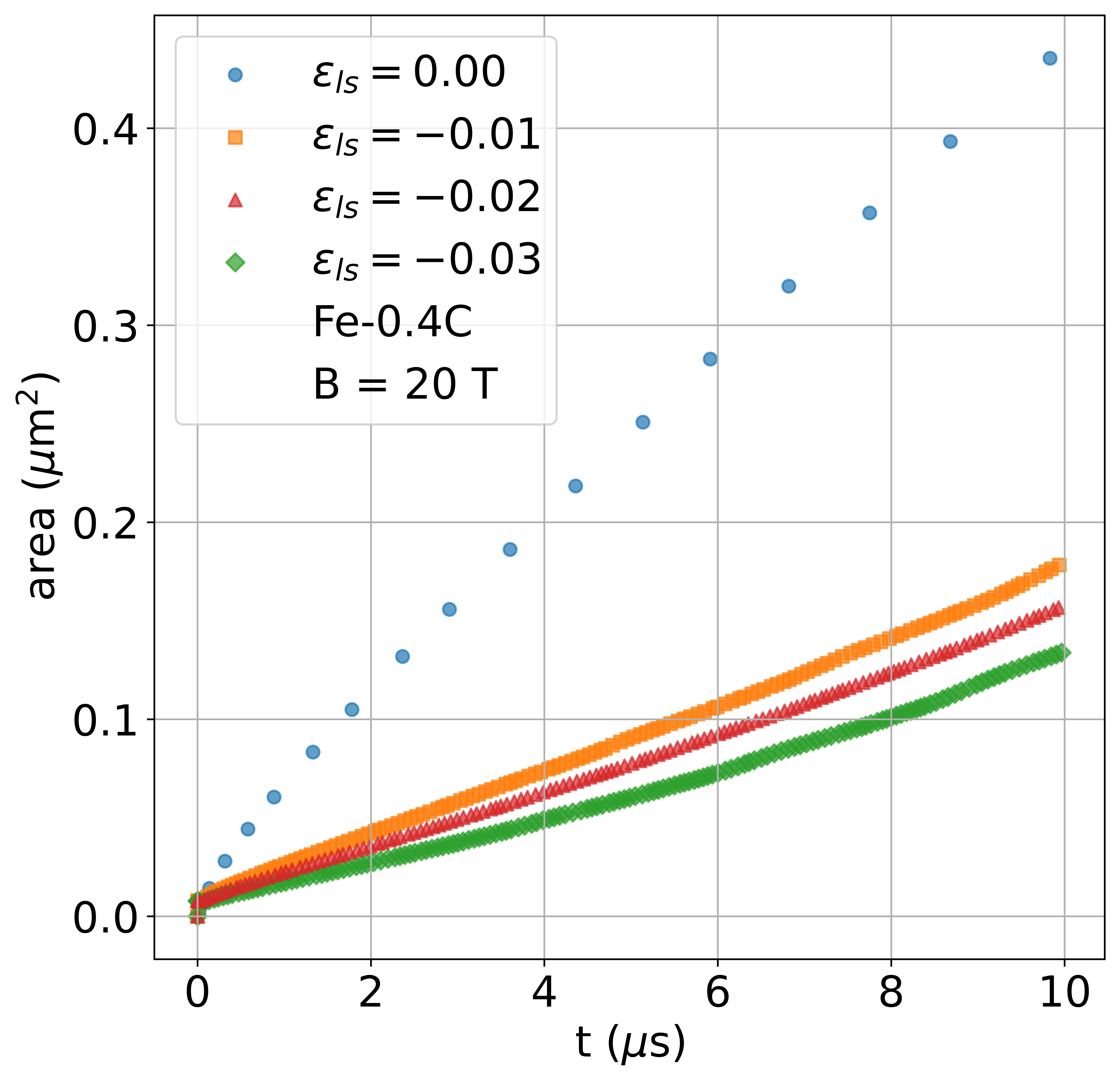}
        \caption{}
    \end{subfigure}
    \caption {Area vs. time ($t$) for the growth of single  precipitates under different conditions: (a) $B=0$ T, $\epsilon_{ls}=0$, and Fe-$x$\%C (with $x = 0.1,\ 0.2,\ 0.3,\ 0.4$); (b) $B=0,\ 1,\ 10,\ 20$ T, $\epsilon_{ls}=0$, for Fe-$0.4$\%C; (c) $B=10$ T, $\epsilon_{ls}=0,\ -0.01,\ -0.02,\ -0.03,\ -0.04$ for Fe-$0.2$\%C; (d) $B=20$ T, $\epsilon_{ls}=0,\ -0.01,\ -0.02,\ -0.03,\ -0.04$, for Fe-$0.4$\%C.}
     \label{fig:grwrate}
\end{figure*}

We also investigate how the composition, eigenstrain, and magnetic field impact the growth rate of the precipitates. Figures~\ref{fig:grwrate} (a)-(d) present the precipitate growth profiles (area vs. time) for various cases. The growth rate of the precipitates tends to be parabolic, such that the area vs. time plot has a linear profile, until the area gets above around
$\SI{0.4}{\micro\meter}^2$. The growth rate slows for larger precipitates. 

Figure~\ref{fig:grwrate}(a) displays the case of purely chemical driving forces at different carbon concentrations. We observe that the growth rate increases with increasing C composition, with the Fe-0.4 wt\%C being the most nonlinear.
Figure~\ref{fig:grwrate}(b) plots the growth behavior of the Fe-0.4 wt\%C system under varying magnetic fields. There is little change in the growth rate going from 0 T to 1 T, but decreases with increasing field strength for $B>1$ T. Although we only show Fe-0.4 wt\%C here, the same trend holds for other compositions.
Figures~\ref{fig:grwrate}(c) and (d) show the growth behavior for Fe-0.2 wt\%C (10 T) and Fe–0.4 wt\%C (20 T), respectively, with misfit strains from $0\%$ to $-3\%$. In both cases, stronger compressive strains reduce the growth rate, though the Fe-0.4 wt\%C (20 T) has a large decrease in the growth rate from 0 to -1\% misfit strain and a much smaller decrease between larger misfit strains.

\begin{figure*}[tbph]        
  \begin{subfigure}[t]{0.49\textwidth}
        \centering
        \includegraphics[width=1.0\linewidth]{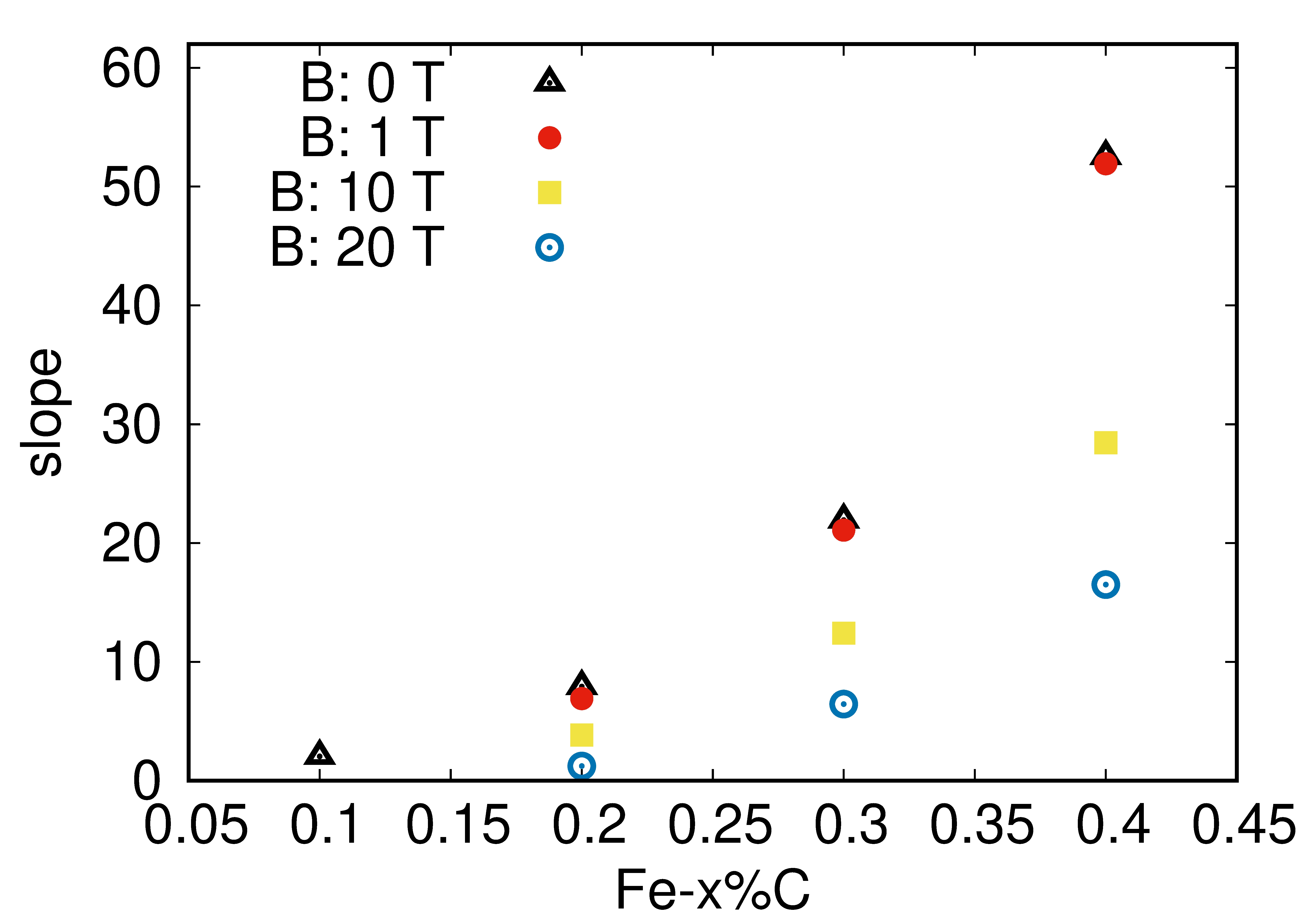}
        \caption{}
        \label{slope1}
    \end{subfigure}
     \begin{subfigure}[t]{0.49\textwidth}
        \centering
        \includegraphics[width=1.0\linewidth]{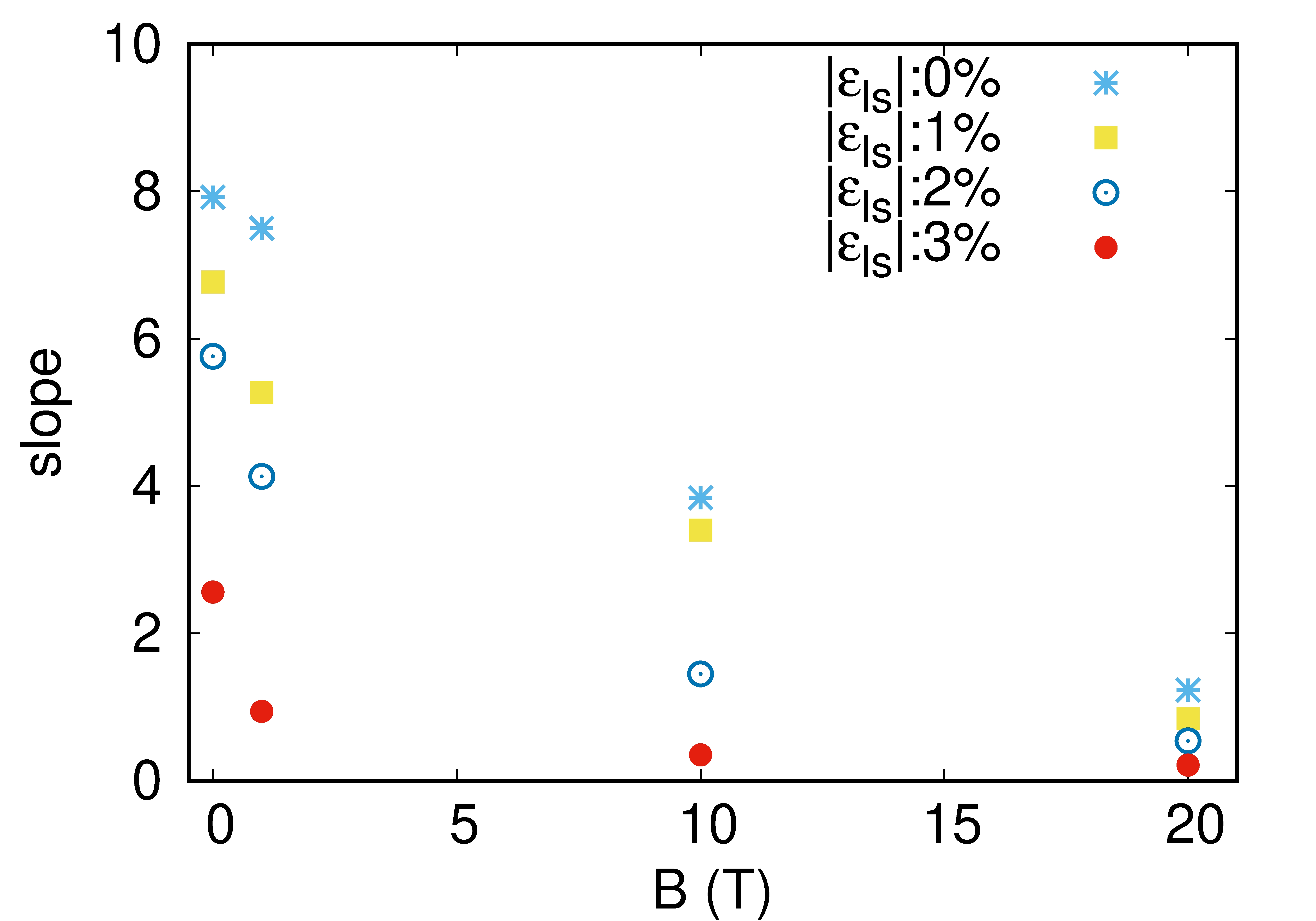}
       \caption{}
       \label{slope2}
    \end{subfigure}
     \begin{subfigure}[t]{0.49\textwidth}
        \centering
        \includegraphics[width=1.0\linewidth]{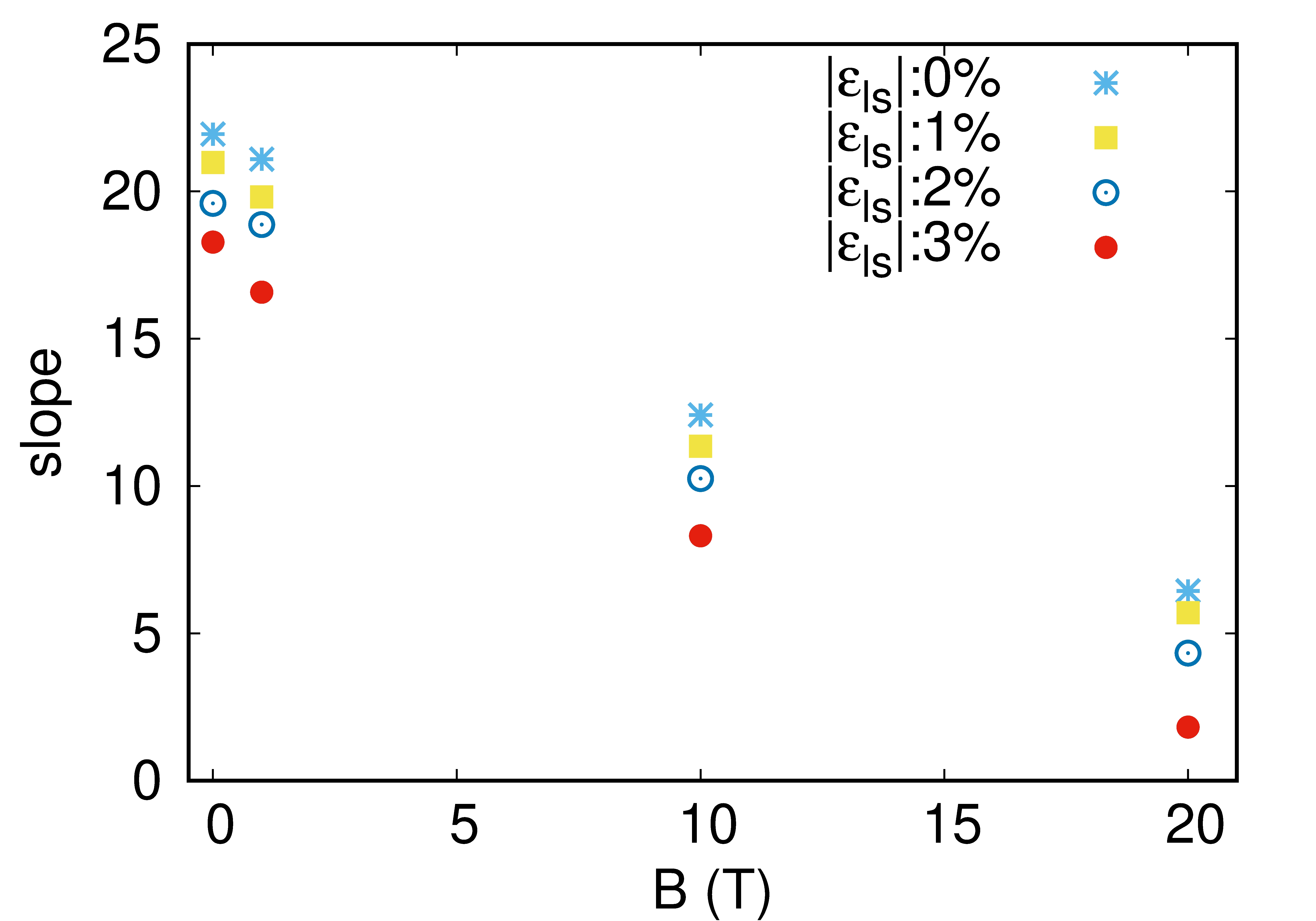}
        \caption{}
        \label{slope3}
    \end{subfigure}
    \begin{subfigure}[t]{0.49\textwidth}
        \centering
        \includegraphics[width=1.0\linewidth]{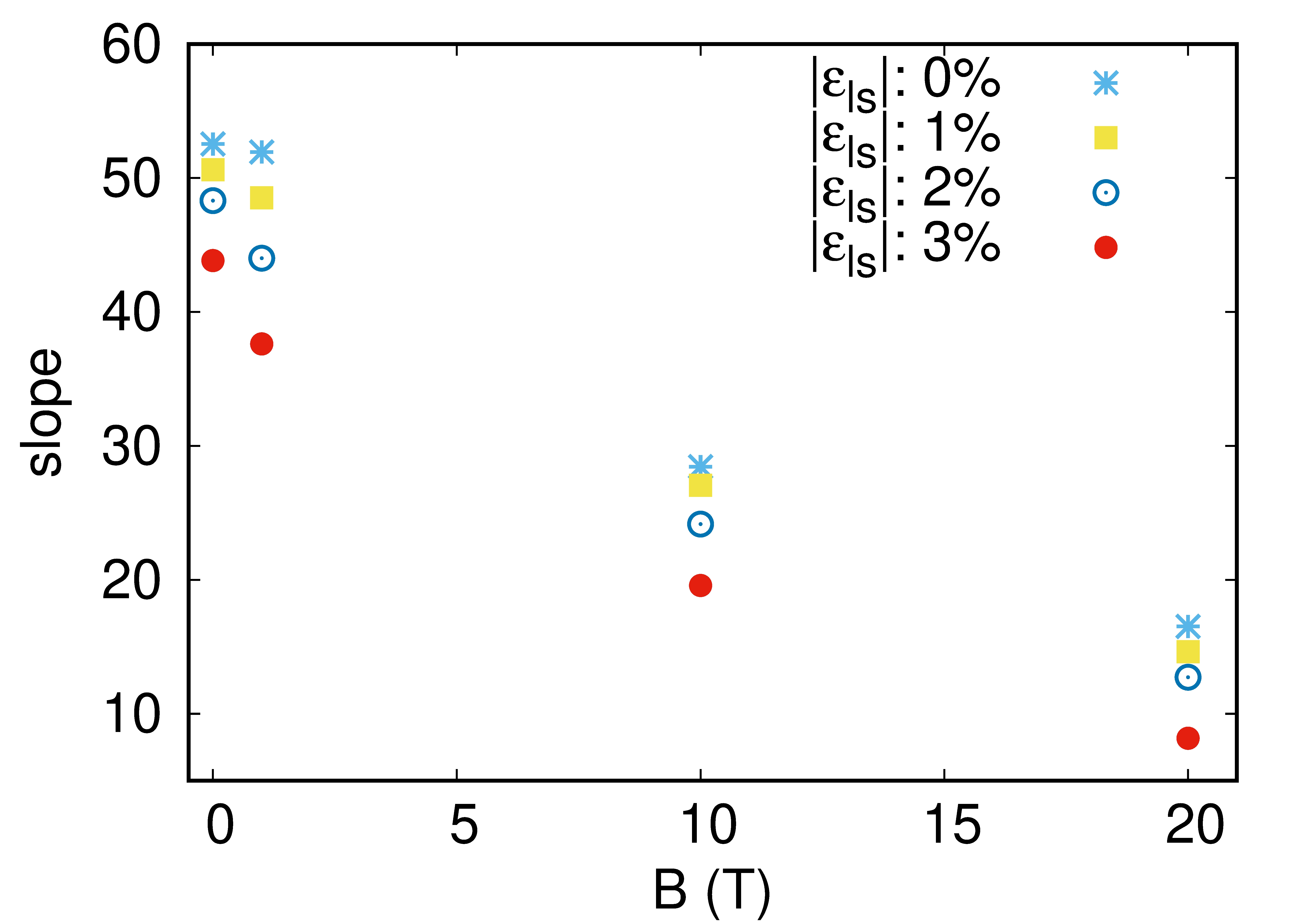}
        \caption{}
        \label{slope4}
    \end{subfigure}

    \caption{Slopes of lines fitted to the area vs. $t$ data shown in Fig.~\ref{fig:grwrate} from the single particle simulations with (a) varying magnetic field and C composition with $\epsilon_{ls}=0$, and (b) - (d) varying magnetic field and misfit strain for Fe-0.2 wt\% C, Fe-0.3 wt\% C, and Fe-0.4 wt\% C, respectively.}
    
    \label{fig:slope}
\end{figure*}

To more quantitatively analyze the growth rates, we fit lines to the area vs.\ time data and the values of the slopes of these lines are compared in Fig.~\ref{fig:slope}. Figure \ref{slope1} shows the growth rate versus C concentration for various field strengths but no lattice strain. The growth rate increases with increasing carbon concentration, but decreases with increasing field strength (though the decrease from 0 T to 1 T is very small). 
Figures~\ref{slope2}-~\ref{slope4} illustrate the variation of the growth rates for different carbon concentrations under combined 
magneto-elastic interactions. Figure~\ref{slope2} displays the growth rates of the Fe-0.2 wt-\%C system subjected to varying magnetic 
field strength and misfit strain, while Figs.~\ref{slope3}-~\ref{slope4} present analogous profiles for Fe-0.3 wt-\%C and 
Fe-0.4 wt\%C systems. We observe a clear trend: increasing magnetic field systematically suppresses the growth rate, while higher 
carbon concentrations promote it. For Fe–0.2 wt\% C (Fig.~\ref{slope2}), the slopes are relatively small even in the absence of strain 
and as the magnetic field increases from 0 to 20 T the growth rate rapidly diminishes, with misfit strains of 2–3\% nearly suppressing the 
growth altogether. Increasing the carbon content to 0.3 wt\% (Fig.~\ref{slope3}) increases the slopes,
indicating the increase in the 
chemical contribution providing stronger thermodynamic driving force for particle growth; however, the retarding effects of both 
magnetic field and strain remain pronounced, leading to reduction in the slope at higher field strength and lattice strain. 
The most pronounced growth is observed for Fe–0.4 wt\% C (Fig.~\ref{slope4}),
where we obtain the highest growth rate under strain-free conditions at low magnetic field demonstrating the dominance of increased chemical driving force in accelerating kinetics. Yet, here too, the combined 
effect of magnetic field and strain substantially hinders the kinetics.

\subsection{Multi-particle system}\label{mulpart}
\begin{figure}[tbp]
\centering
  \includegraphics[width=1.0
  \linewidth]{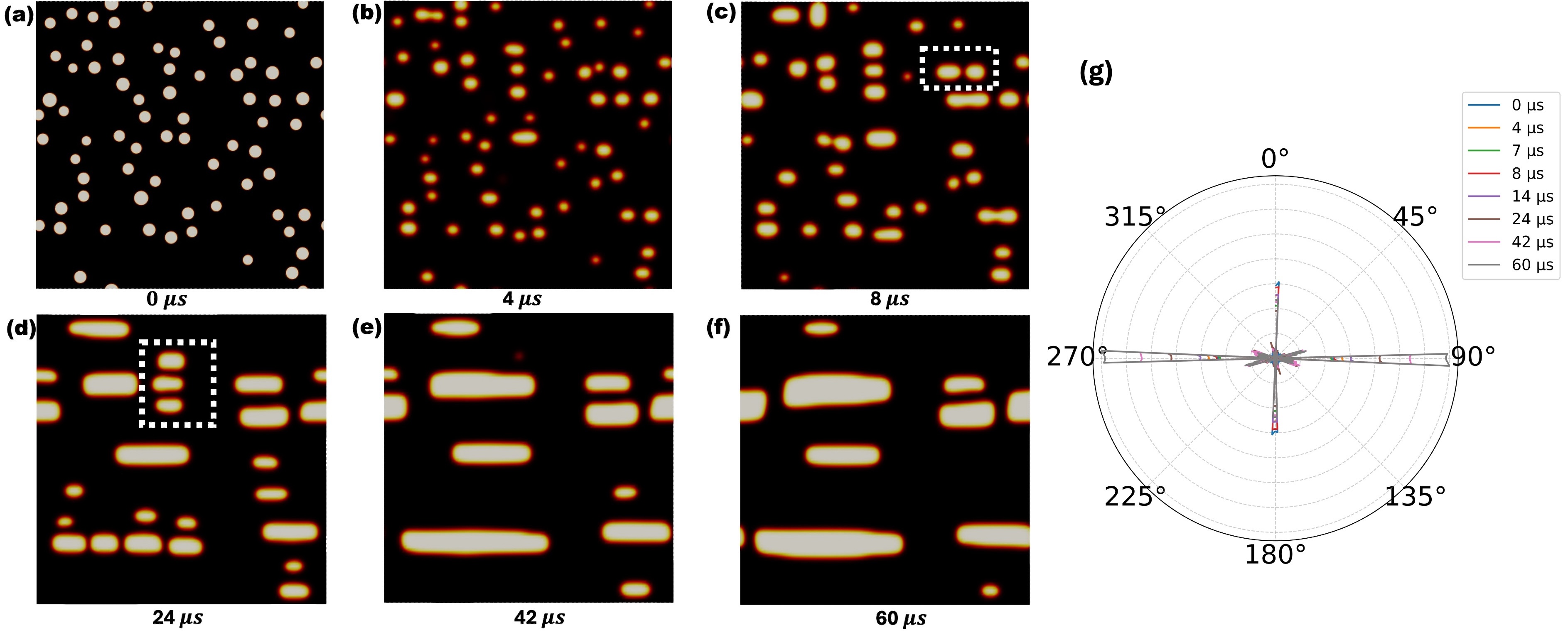}
  \caption{Time‐lapse evolution of the Fe-0.4wt\%C multi‐particle simulation in a $\SI{1}{\micro\meter} \times \SI{1}{\micro\meter}$ domain with 70 initial particles with a 20 T field and $\epsilon_{ls} = -0.03$ at: (a) \SI{0}{\micro\second}, (b) \SI{4}{\micro\second},
  (c) \SI{8}{\micro\second}, (d) \SI{24}{\micro\second},
  (e) \SI{42}{\micro\second}, and (f)\SI{60}{\micro\second}. (g) Temporal evolution of the power spectrum for the multi-particle system, illustrating the development of anisotropy and alignment over time.
  }
\label{fig:combr2}
\end{figure}

After studying the coupled chemo-magneto-mechanical effects on the microstructural characteristics in single precipitate morphology, we further extend our model to investigate the combined field effects on systems with multiple particles. To perform the multi-particle studies we consider a simulation domain size of $\SI{1}{\micro\meter} \times \SI{1}{\micro\meter}$ with 70
circular particles with an average radius of \SI{20}{\nano\meter}. The particles are distributed randomly through the domain and their radii also vary randomly according to a normal distribution with a standard deviation of \SI{2}{\nano\meter}. 
A minimum center to center distance of \SI{60}{\nano\meter} was maintained between the particles such that initially they do not overlap.
Figure~\ref{fig:combr2}(a) shows the initial  microstructure. We model the microstructure evolution of a Fe-0.4wt\%C alloy (supersaturation value $c_\infty = 0.0182$ mole fraction) with a magnetic field of $20~\textrm{T}$ and a misfit strain of $-3\%$. 

As the microstructure evolves (Figs.~\ref{fig:combr2}(b)–(f)), the precipitates adopt elongated, brick-like shapes, consistent with the morphologies observed in similar single particle simulations.  To quantify the elongation, we use the temporal evolution of the angular power, as shown in Fig.~\ref{fig:combr2}(g). The continuous increment of the profiles along  $90\degree$ and $270\degree$ demonstrates the evolution and characteristics of the precipitate alignment along the direction of the applied field.

\begin{figure}[tbp]
\centering
  \includegraphics[width=1.0
  \linewidth]{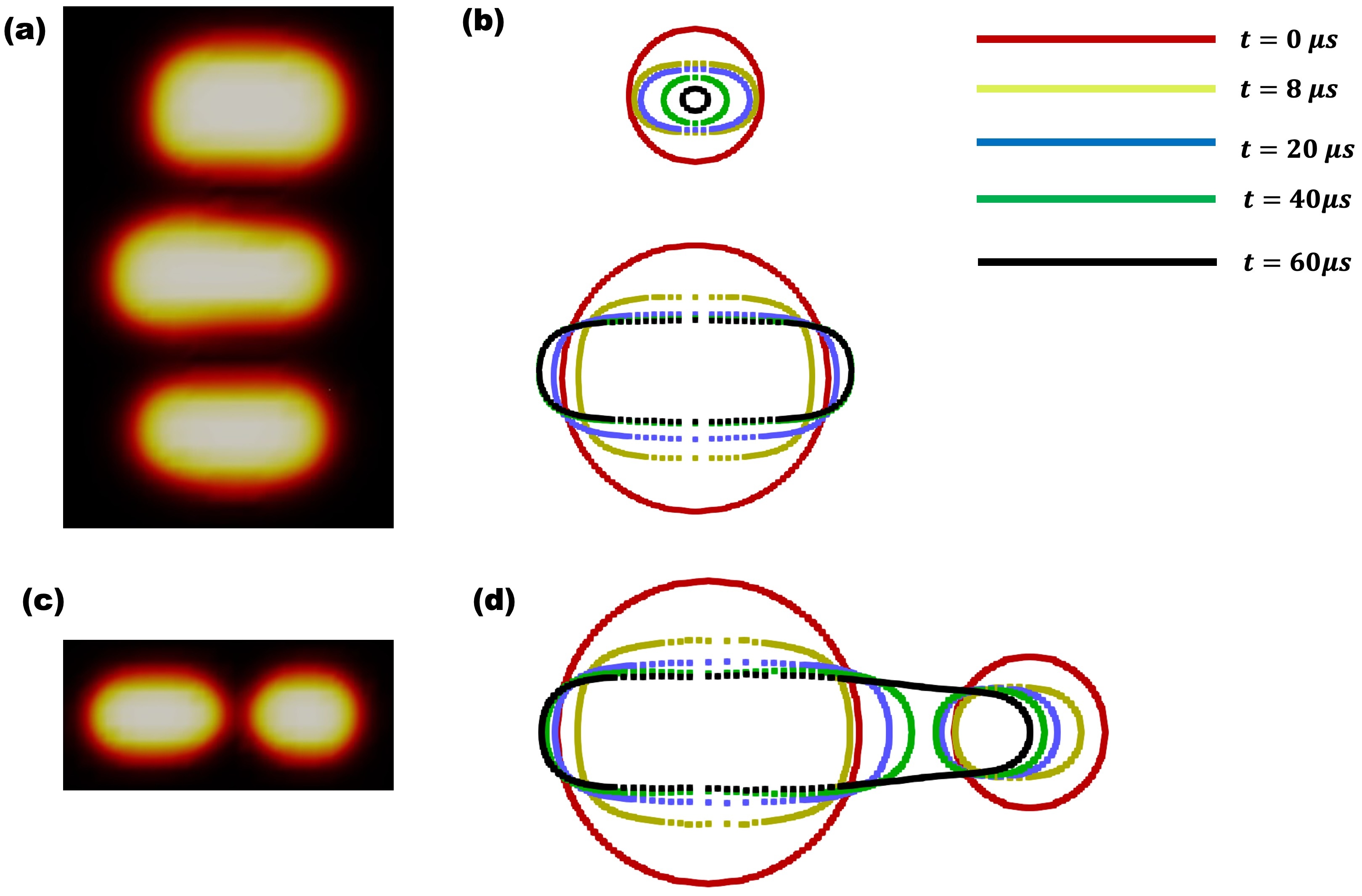}
  \caption{Interactions between particles orientated parallel and perpendicular to the applied field. (a) Example of particles aligned perpendicular to the field from Fig.~\ref{fig:combr2}(d), and (b) contour‐plot evolution of a two-particle simulation with particles oriented perpendicular to the field. (c)  Example of particles aligned parallel to the field from Fig.~\ref{fig:combr2}(c); (d) contour‐plot evolution of a two-particle simulation with particles oriented parallel to the field.}
\label{fig:combr3}
\end{figure}

Furthermore, in this simulation there are two general types of particle interactions taking place. While, the particles placed horizontally tend to align 
and coalesce in the direction of the applied field (see Fig.~\ref{fig:combr2}(c)), they tend to coarsen
Fig.~\ref{fig:combr3}(c) when placed vertically or perpendicular to the applied field (see Fig.~\ref{fig:combr2}
(d) and Fig.~\ref{fig:combr3}(a)). To explore these interactions in more detail, we use specific two-particle simulations. First, 
we simulate the growth of two circular particles of different radii that are aligned perpendicular to the field.  The 
bigger particle has an initial radius of \SI{40}{\nano\meter} and the smaller one of \SI{25}{\nano\meter}, as shown in 
Fig.~\ref{fig:combr3}(b). A typical coarsening behavior is observed, as the smaller particle shrinks and the larger 
particle grows. Second, we simulate two circular particles of the same size but that are aligned parallel to the field, as shown in Fig.~\ref{fig:combr3}(d). The two particles attract each other 
and undergo directional coalescence. 

\subsection{Magnetic field-dependent Diffusivity}
All the simulations performed in the previous sections assume that the C diffusivity is independent of magnetic field strength. However, studies indicate that an external magnetic field significantly reduces the C diffusivity \cite{fujii2011diffusion,liu2011effects}. To capture the magnetic field effects on the diffusivity, we introduce a magnetic‐energy contribution into the activation barrier \cite{liu2011effects}:
\begin{equation}
    Q_{eff} = Q_{i} -\alpha_B B^2, 
\end{equation}
where $Q_i$ defines the activation energy of the system, and $\alpha_B$ (in J/T$^2$mol) is a parameter obtained by fitting to experiments \cite{liu2011effects}.   
Thus, the diffusivity with the magnetic field effect $D_m$ can be written as a function of magnetic field as:
\begin{equation}
\begin{split}
    D_m(B,T) &= D_{0}e^{-Q_{eff}/RT},\\   
         & = D_{0}e^{-(Q_i - \alpha_B B^2)/RT},\\
         & = De^{\alpha_B B^2/RT}.
 \end{split}
 \label{effecD}
\end{equation}

\begin{figure*}[tbp]
    \begin{subfigure}[t]{0.49\textwidth}
        \centering
        \includegraphics[width=1.0\linewidth]{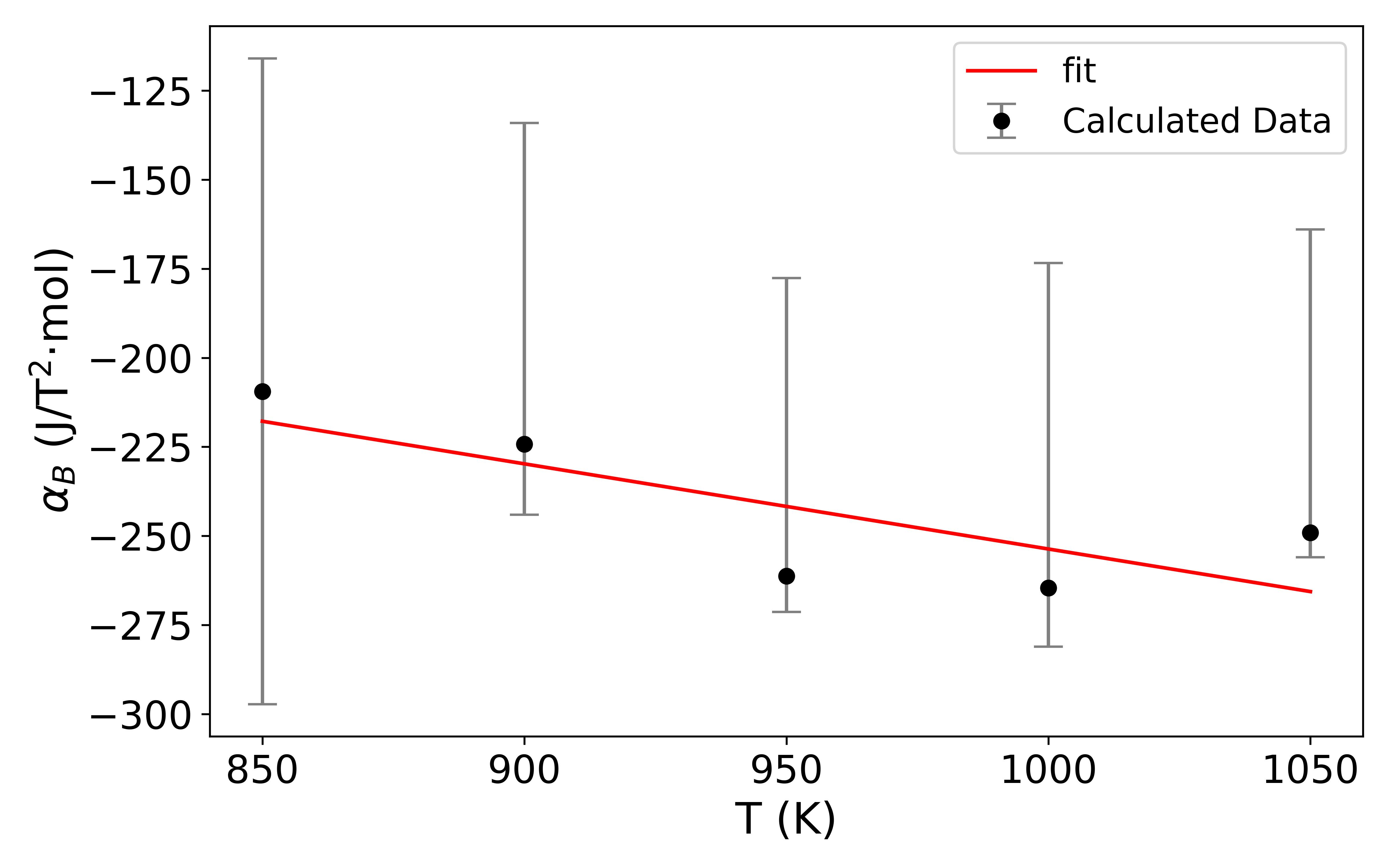}
        \caption{}\label{fig:fitting}
    \end{subfigure}
    \begin{subfigure}[t]{0.57\textwidth}
        \centering
        \includegraphics[width=1.0\linewidth]{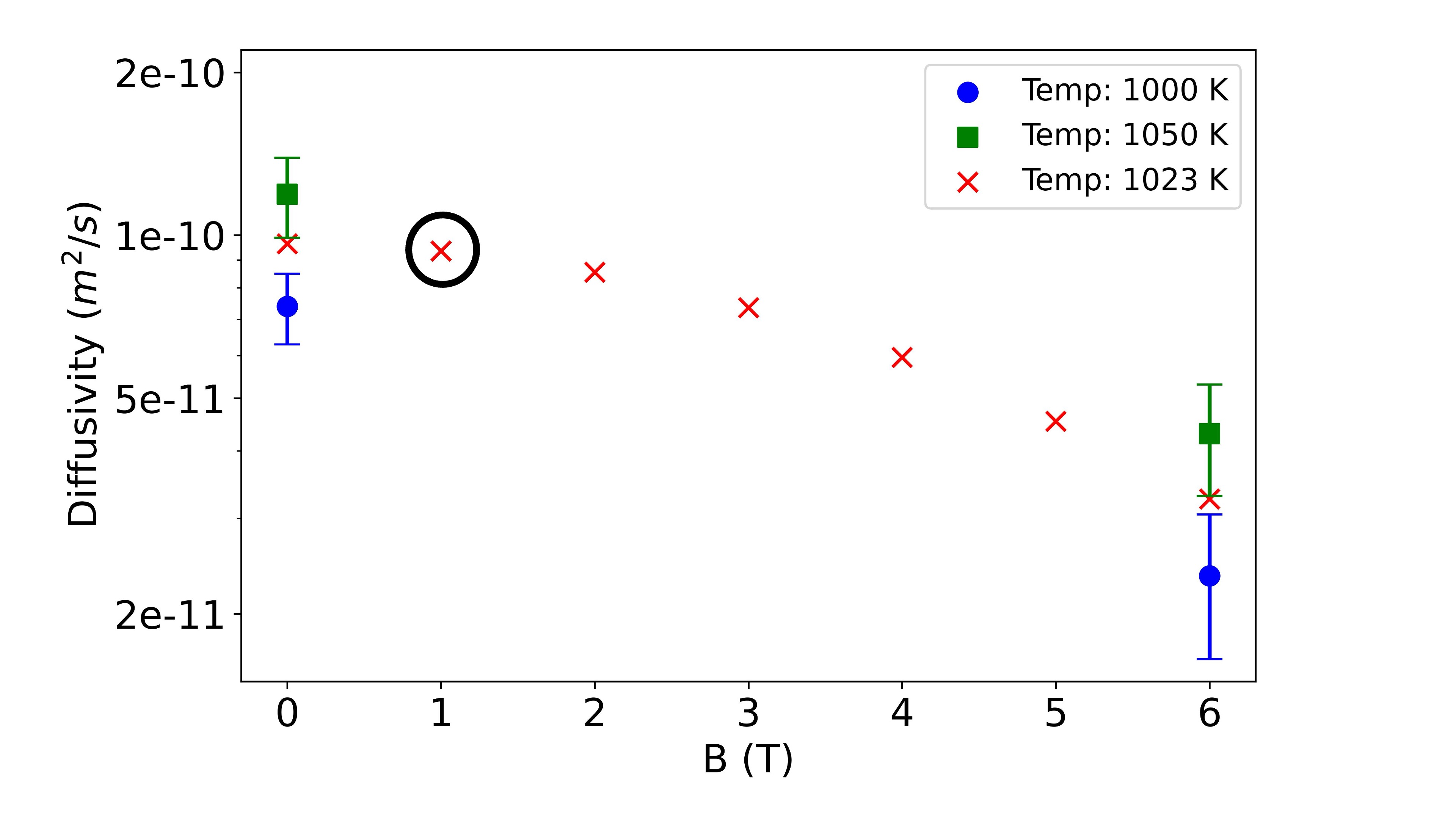}
        \caption{}\label{diff_coef}
    \end{subfigure}
    
     \begin{subfigure}[t]{0.49\textwidth}
        \centering
        \includegraphics[width=0.8\linewidth]{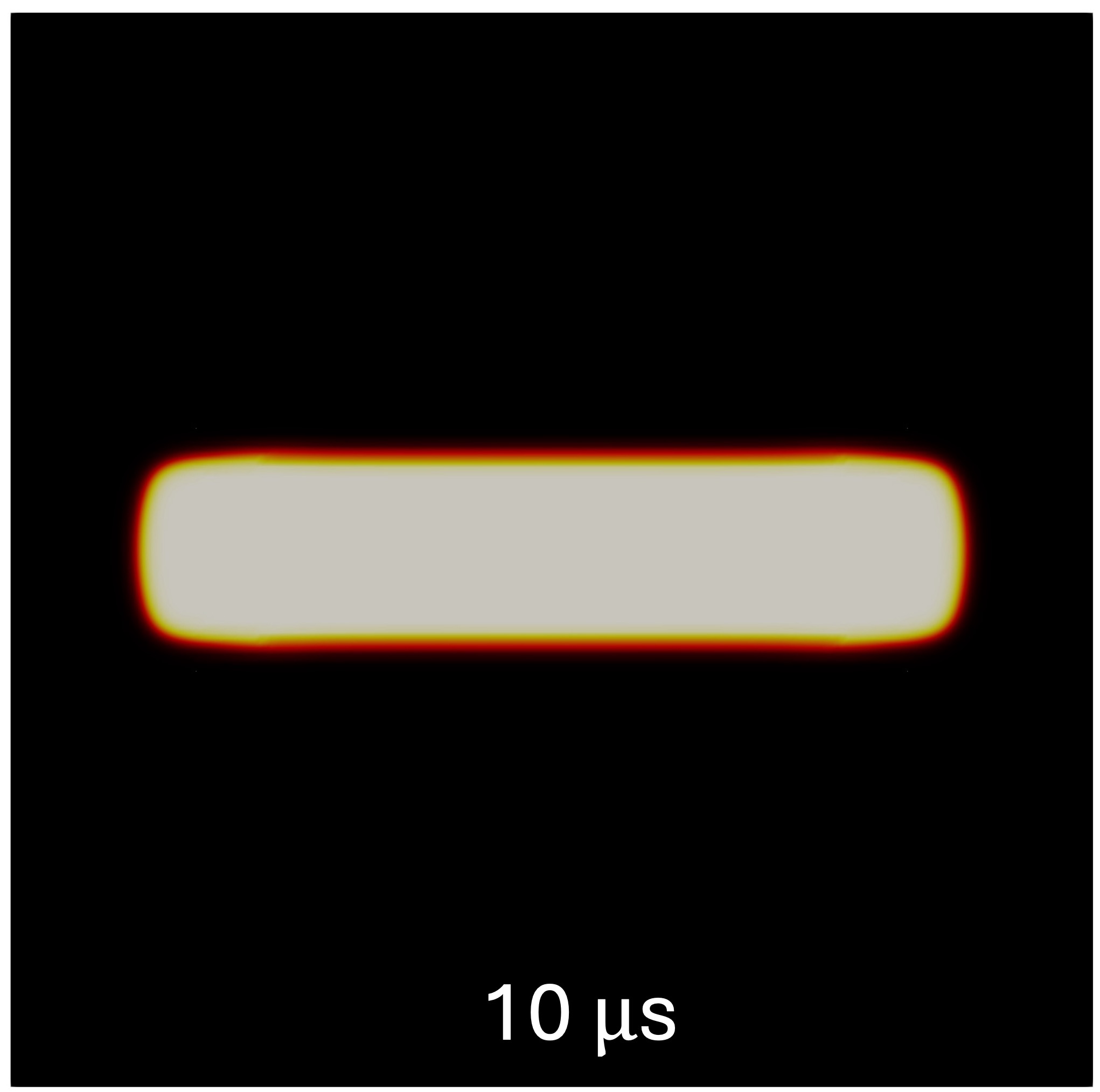}
        \caption{}\label{magdiff1}
    \end{subfigure}
    \begin{subfigure}[t]{0.49\textwidth}
        \centering
        \includegraphics[width=0.8\linewidth]{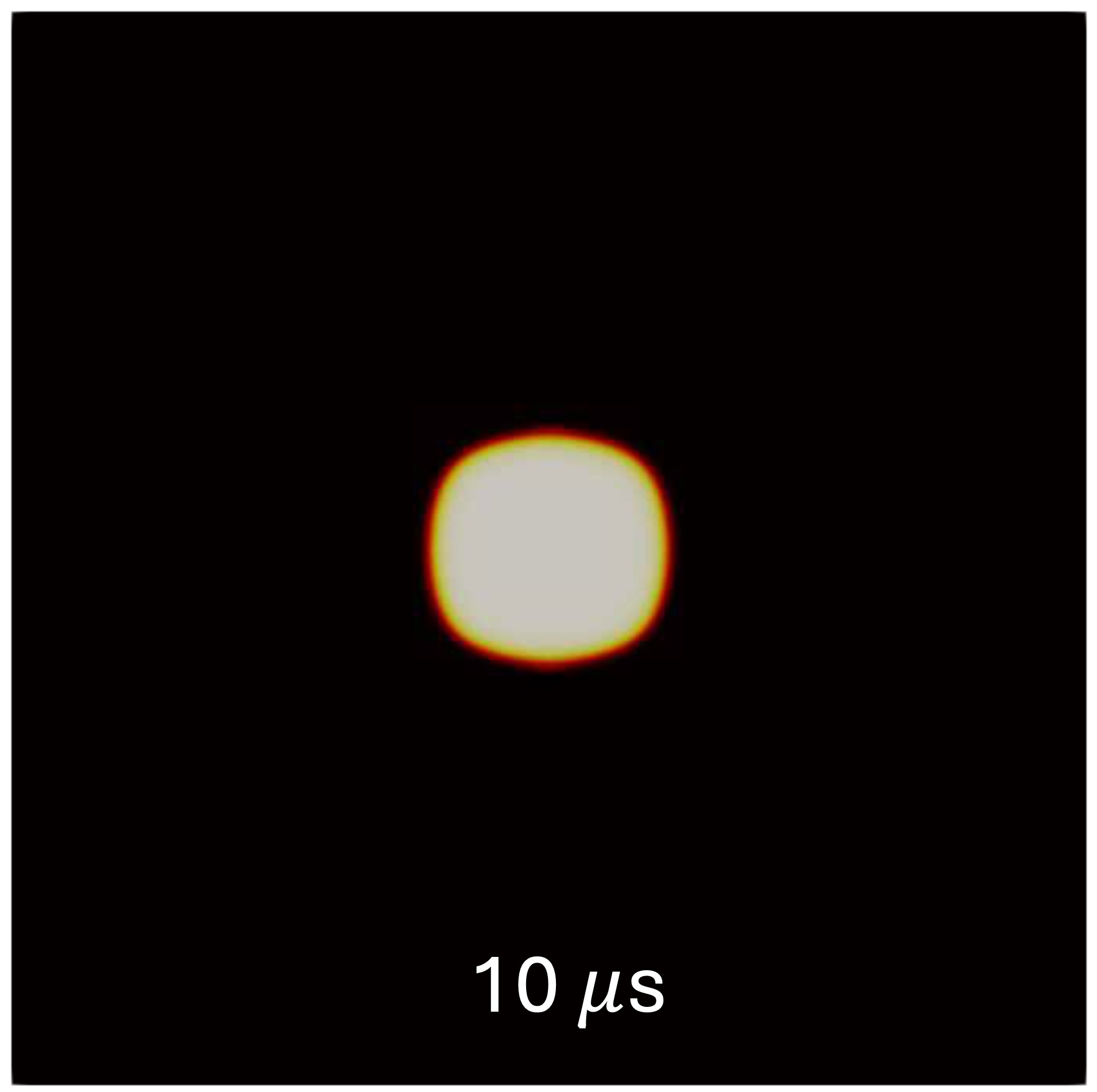}
        \caption{}\label{magdiff2}
    \end{subfigure}

     \begin{subfigure}[t]{0.49\textwidth}
        \centering
        \includegraphics[width=0.8\linewidth]{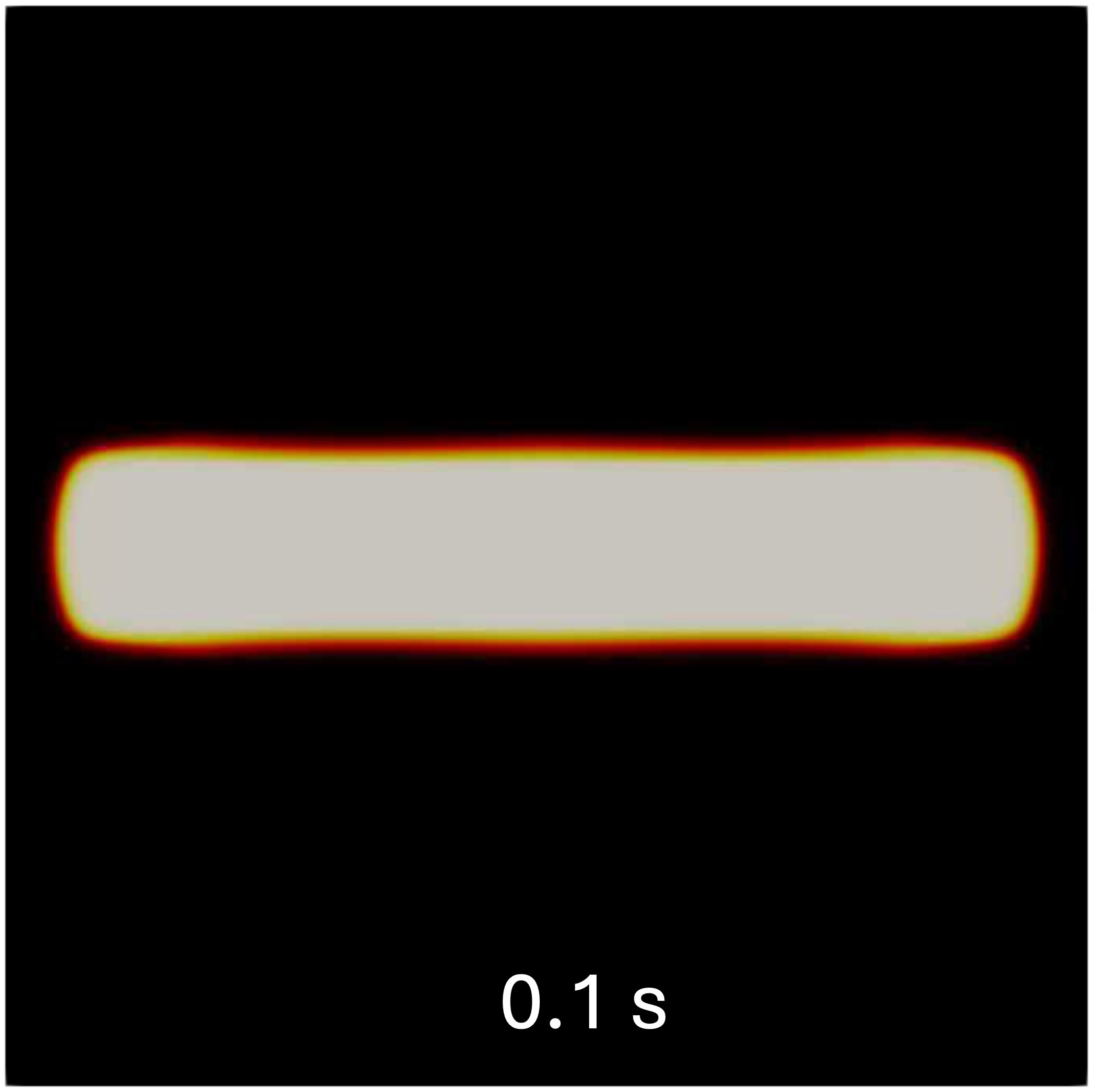}
        \caption{}\label{magdiff3}
    \end{subfigure}
    \begin{subfigure}[t]{0.49\textwidth}
        \centering
        \includegraphics[width=1.0\linewidth]{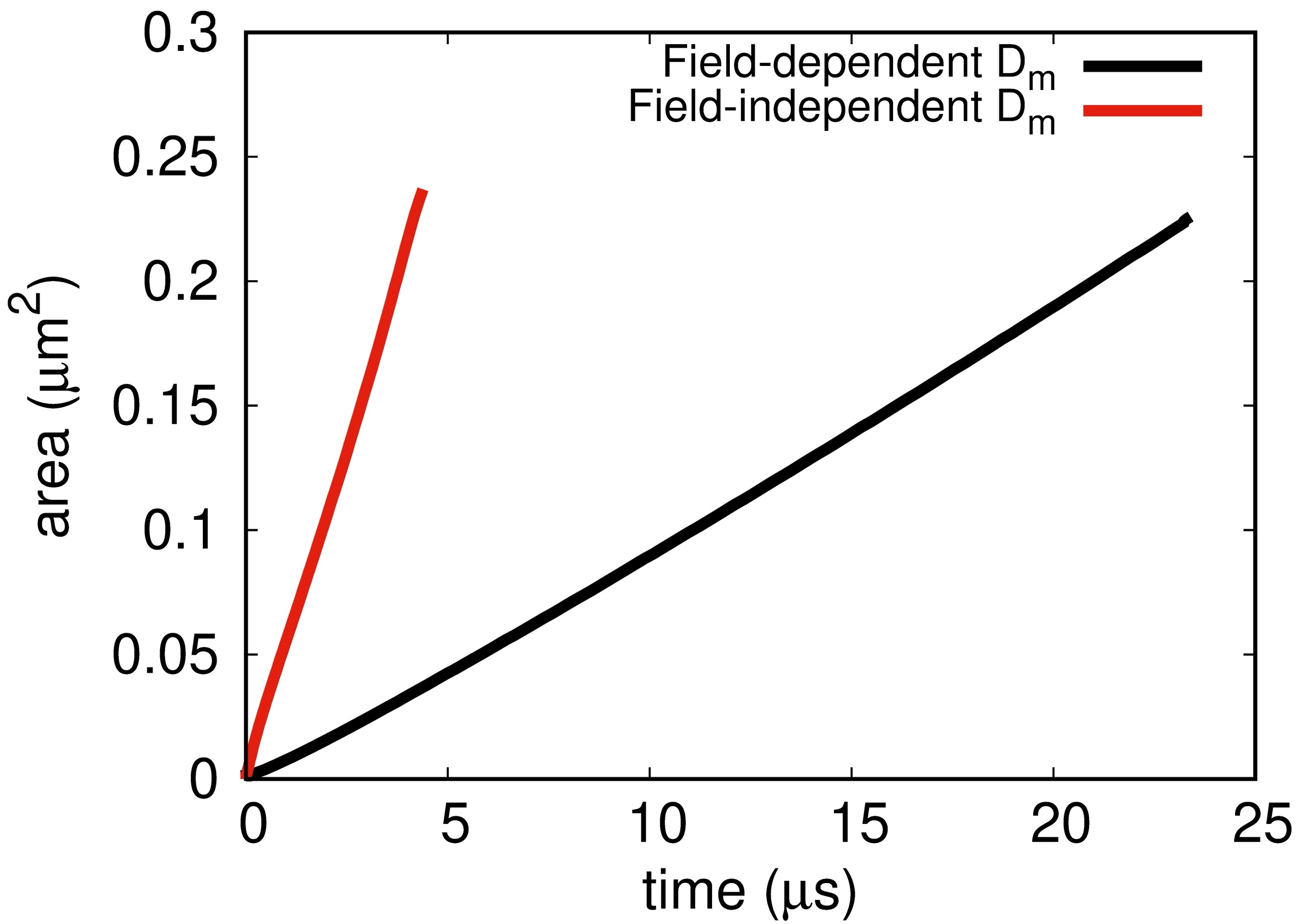}
        \caption{}\label{magdiff4}
    \end{subfigure}
    
    \caption{Fitting process to obtain the C diffusivity as a function of temperature and magnetic field, where (a) shows the fitted line for $\alpha_B$ as a function of temperature for use in Eq.~\eqref{effecD}, along with the data \cite{fujii2011diffusion} (error bars as reported in the paper), and (b) shows the diffusivities at 1023 K estimated from Eq.~\eqref{effecD} using the fit shown in (a), compared to the experimental data at 1000 and 1050 K. The field dependent diffusivity data at 1 T (marked by black circular region) is used to perform the phase field simulation. The impact of magnetic field-dependent diffusivity on the growth of a single particle with $T=1023$ K, for Fe-0.4\%C, $\epsilon_{ls} = -0.03$, and $B^{ext} = 1$ T. The morphology after $\SI{10}{\micro\second}$ is shown for the (c) normal diffusivity and (d) magnetic field-dependent diffusivity. (e) Morphology after 0.1 s  with a magnetic field-dependent diffusivity.
    (f) Area vs. time profile with and without the impact of magnetic field on the diffusivity}
    \label{magdiff}
\end{figure*}

To determine a value for $\alpha_B$, we fit the diffusivity to values measured in the  literature~\cite{fujii2011diffusion}.
We obtain an expression for $\alpha_B$ by rewriting Eq.~\ref{effecD} as 
\begin{equation}
    \alpha_B = \frac{RT}{B^2}ln\left(\frac{D_m(B,T)}{D}\right),
    \label{alphab}
\end{equation}
where the values of $D_m$ for various applied fields and temperatures are obtained from experiments. Studies report the magnetic field-dependent diffusivity data for different temperatures ranging from $\SI{850}{\kelvin}$ to $\SI{1250}{\kelvin}$~\cite{fujii2011diffusion}. However, we do not find any data for the temperature of $\SI{1023}{\kelvin}$ used in this study.
In order to obtain the values of $\alpha_B$ from the diffusivity data, as reported by Fujii et al. 
\cite{fujii2011diffusion}, we fit a line to the data  as a function of temperature, obtaining $\alpha_B = -0.2391 T - 14.6$. 
The values of $\alpha_B$ along with the fitted profile is shown in Fig.~\ref{fig:fitting}.

Having fit an expression for $\alpha_B$ as a function of temperature to the data from \cite{fujii2011diffusion}, we can use Eq.~\eqref{effecD} to estimate the C diffusivity at  $\SI{1023}{\kelvin}$ for an applied magnetic field up to the 6 T field used in the experiments. Estimating the diffusivity for higher fields would be extrapolating outside the data range and is not recommended. Figure~\ref{diff_coef} shows the estimated values at 1023 K, as well as experimental values from \cite{fujii2011diffusion} for 1000 K and 1050 K. The estimated diffusivity at 1023 K at 0 T and 6 T falls between the experimental values from 1000 K and 1050 K, as shown in Fig.~\ref{diff_coef}.

 
After obtaining the diffusivity data as a function of magnetic fields, we now run a single particle simulation similar to those from Section \ref{sec:results_single} at 1023 K for Fe-0.4\%C ($c_\infty = 0.0182$), $\epsilon_{ls} = -0.03$, and $B^{ext} = 1$ T, but including the impact of the applied field on the diffusivity. We also compare the particle growth behavior with and without the magnetic field effect on the diffusivity. Figure~\ref{magdiff1} shows the microstructure obtained after $\SI{10}{\micro\second}$ using the original, constant diffusivity, and
 Fig.~\ref{magdiff2} with the magnetic field-dependent diffusivity (diffusivity data is obtained from Eq.~\eqref{effecD} and displayed by the marked circular region in Fig~\ref{diff_coef}).
Although both particles have grown for the same time, the particle is much smaller when we consider the smaller diffusivity due to the applied field (Fig.~\ref{magdiff2}. Due to the slower kinetics, it has not had time to elongate like in the case where we ignore the impact of field on the diffusivity. If the simulation is run much longer ($\SI{0.1}{\second}$), the field‐dependent case does elongate, as shown in Fig.~\ref{magdiff3}.
The particle area vs.\ time with and without the impact of the magnetic field on the diffusivity is shown in Fig.~\ref{magdiff4}. The field‐dependent case exhibits much slower kinetics, consistent with its lower diffusivity.

Up to now, we have assumed that the impact of the magnetic field on the diffusivity is isotropic. However, it is often hypothesized that in ferromagnetic materials such as Ferrite ($\alpha$), magnetostriction will create direction-dependent lattice strains that modify interstitial atom migration barriers~\cite{souissi2016ab}, breaking the rotational symmetry. This broken symmetry leads to anisotropic diffusivity and mobility~\cite{wang2011effect}. However, anisotropy in the magnetic field effect on the diffusivity has not been measured experimentally. Here we derive an expression for the  anisotropic mobility due to the magnetic field, suitable for integration into our phase‐field simulation model.

According to the Jiles-Atherton (J-A) framework, we can approximate the magnetostriction in a cubic Fe system as~\cite{jiles2002theory}:
\begin{equation}
    \lambda(\mathbf{B}) \approx \lambda_{100}[1 - e^{-\mathbf{B}/B_0}]^2,
    \label{magneto}
\end{equation}
where $\lambda_{100}$ refers to the saturation magnetostriction and $B_0$ is the Jiles-Atherton shape parameter. Assuming an applied uniaxial magnetic field along the $x$-axis, magnetostriction will produce a uniaxial strain in the $x$-direction and a transverse strain (due to Poisson Effect) in the $y$–$z$ plane~\cite{cullity2011introduction}. Thus, for bcc-Ferrite ($\alpha$) under a strong magnetic field the axial strain $\epsilon_\parallel^{mstr} = \lambda_{100} = \num{20e-06}$ and  transverse strain $\epsilon_\perp^{mstr} = -\nu\lambda_{100} = \num{6e-06}$. 

The anisotropic diffusivity is a Rank 2 diagonal tensor. In this case, the anisotropy comes from the different shifts in the migration barrier due to the lattice strains induced by the magnetic field. The shift in the migration barrier under magnetic field can be rewritten using a Taylor expansion as (approximated to first order):
\begin{equation}
    Q( \boldsymbol{\epsilon}^{mstr}) = Q_0 + \frac{\partial Q}{\partial \boldsymbol{\epsilon}^{mstr}}\bigg{|}_{\boldsymbol{\epsilon}^{mstr}=0}\boldsymbol{\epsilon}^{mstr} + \mathcal {O}({\boldsymbol{\epsilon}^{mstr}}^2),
\end{equation}
where $Q_0$ is the unstrained barrier height. If we assume  $\mathbf{K} =  \frac{\partial Q}{\partial\boldsymbol{\epsilon}^{mstr}}$ 
the shift in the barrier can be defined as:
\begin{equation}
   \begin{split}
    \Delta Q &= Q(\boldsymbol{\epsilon}^{mstr}) - Q_0 = \mathbf{K}\cdot\boldsymbol{\epsilon}^{mstr}\\
             &=k_{xx}\epsilon_{xx}^{mstr} + k_{yy}\epsilon_{yy}^{mstr}\\
             &= Q_{\parallel} + Q_{\perp} = Q_i,
    \label{barrier}
\end{split}
\end{equation}
where $Q_i$ refers to energy barrier in the direction parallel ($\parallel$) or perpendicular ($\perp$) to the magnetic field. 
Studies based on first principle density functional theory report that the  energy changes by roughly $10$ eV/atom $\approx \num{9.7e05}$ J/mol per unit strain~\cite{souissi2016ab,domain2004ab}. Thus, the barrier height can be rewritten as:
 \begin{equation}
 \begin{split}
     & Q_\parallel = k\epsilon_\parallel^{mstr} \approx 19~\textrm{J/mol},\\
     & Q_\perp \approx -5.8~\textrm{J/mol}.
 \end{split}
 \end{equation}
Hence, the diagonal values of the anisotropic diffusivity $\mathbf{D}$ takes the form:
\begin{equation}
    D_{ii} = D^{iso} e^{-Q_i/RT}, 
    \label{eq:aniso_D}
\end{equation}
where $D^{iso} = D_{m}(\mathbf{B}, T)$ is the magnetic-field dependent isotropic diffusivity as described in Eq.~\eqref{effecD}. Thus, using Eq.~\eqref{eq:aniso_D}, we rewrite the diffusivities along the parallel and perpendicular to the magnetic field direction as:
\begin{subequations}
  \begin{align}
 D_{xx} &=D_{\parallel}= D^{iso} e^{-Q_{\parallel}/RT}, \\
 D_{yy} &=D_{\perp}= D^{iso} e^{-Q_{\perp}/RT}.
  \end{align}
   \label{eq:aniso_Ddirec}
\end{subequations}

Based on this model, we can estimate the different diagonal components of the anisotropic diffusivity for several applied fields, as shown in Table~\ref{table3}. 
\begin{table}[bthp]
\begin{center}
\caption{Anisotropic diffusivity components calculated using Eq.~\eqref{eq:aniso_D}.}
\label{table3}
\begin{tabular}{||c|c|c|c||} 
 \hline
 Magnetic field $\mathbf{B}$ (T) & $D_\perp(\mathbf{B})$ ($m^2/s$) & $D_\parallel(\mathbf{B})$ ($m^2/s$) \\ [0.5ex] 
 \hline\hline
  1 & \num{9.37e-11} & \num{9.35e-11} \\ 
  \hline
   2 & \num{8.56e-11} & \num{8.54e-11} \\
   \hline
    3 & \num{7.36e-11} & \num{7.34e-11} \\
   \hline
     4 & \num{5.95e-11} & \num{5.94e-11} \\
   \hline
    5 & \num{4.54e-11} & \num{4.53e-11} \\
   \hline
    6 & \num{3.26e-11} & \num{3.25e-11} \\
   \hline
   10 & \num{4.73e-12} & \num{4.72e-12} \\ 
\hline
\end{tabular}
\end{center}
\end{table}

For all field strengths, the two components only vary slightly ($<2\%$). 
Given the small magnitude of this anisotropy, it is unlikely to have a major impact on the overall microstructural evolution in diffusion-controlled system
since the chemical, interfacial, and elastic energies and dipolar alignment dominate.


\section{Discussion} \label {sec:discussion}
In this study, we focus on delineating the physics governing the formation of elongated microstructures in 
thermo-magnetically processed steels. In this implementation, we have introduced some physics-based assumptions, 
particularly in the treatment of magnetic interactions. Because the easy and the hard magnetization direction of the 
$\alpha$ is indistinguishable at the considered scale, we have neglected the effects of intrinsic magnetocrystalline 
anisotropy~\cite{zeng2022gibbs}. 
Furthermore, magnetization was represented as a scalar quantity, which precludes the explicit evolution and 
relaxation of the magnetic moments as vectors. This simplification, however, offers the practical advantage of avoiding the 
need to solve the full Landau-Lifshitz-Gilbert (LLG) equation~\cite{landau1992theory,gilbert2004phenomenological, zhang2005phase} while still capturing the essential magnetostatic behavior 
relevant to the microstructural evolution. 

While the anisotropic contribution of the system is negligible, the $\gamma/\alpha$ phase discontinuity creates mesoscopic 
magnetostatic anisotropy that emerges due to non-uniform magnetic fields and surface magnetic charges 
near the phase boundaries. These surface magnetic charges manifest a strong demagnetization field and corresponding 
directional driving force that can promote morphological elongation~\cite{hubert1998magnetic,skomski2008simple}.
As the present formulation employs magnetization as a scalar field, it resolves only the isotropic component of the 
magnetostatic interaction. To compensate for the unresolved vectorial 
and interfacial magnetic contributions, we introduce an anisotropic scaling factor exclusively to the demagnetization part 
to preserve the correct energetic scale predicted by the classical demagnetization theory~\cite{osborn1945demagnetizing} for 
these systems.
Despite these simplifying assumptions, the proposed model is robust enough to capture the essential magneto-mechanical 
coupling and reproduce the experimentally observed field-induced elongation behavior~\cite{maruta2002magnetic, shimotomai2000aligned, ohtsuka2000alignment, shimotomai2003formation}, establishing a physically consistent 
framework for studying magnetic field-assisted microstructural evolution.

Finally, we use 2D simulations in order to reduce the computational cost. This essentially assumes that the growing precipitates are unchanging in the third spatial dimension, which does not accurately represent precipitate morphology. However, since we have applied the magnetic field such that the precipitate elongates in the x-direction, we are able to investigate the causes of the elongated morphology with the 2D simulations. We suggest that future simulations be carried out in 3D to confirm our findings.

Now, we delve deep into analyzing our results from Section~\ref{sec:results}) to comprehend the underlying mechanisms governing the observed microstructural characteristics. For the single particle system shown in Fig.~\ref{fig:evol},
minimization of chemical energy favors a sphere (Fig.~\ref{fig:evol}(b)), whereas minimizing demagnetizing energy favors an elongated ellipsoid (Figs.~\ref{fig:evol}(c)-(e)). The resulting ellipsoidal morphology thus represents a compromise between these two competing energetic contributions. Although an external magnetic field facilitates the alignment of magnetic domains, the elongation itself primarily arises from the internal demagnetization effects at the 
$\gamma/\alpha$ interface.

This behavior can be rationalized by treating the paramagnetic fcc nucleus (volume $V$) embedded in the ferromagnetic $\alpha$ phase (susceptibility $\chi$) as a “magnetic hole” with moment $m = -MV$~\cite{shimotomai2000aligned},
where $M$ is the magnetization induced in the $\alpha$ phase by an external field.
Moreover, the observed elongation of the precipitate along the direction of the magnetic field
indicates strong directional dependence of the local magnetic field. 
This results in an anisotropic
contribution to the driving force that originates from spatial variations in the interfacial magnetic field, leading to direction-dependent driving forces and growth kinetics.

The emergence of such anisotropic interfacial effects is further corroborated by the distinct peaks observed in the FFT power spectrum along $90^\circ$ and $270^\circ$ (Fig.~\ref{fig:fft}(g)-(l)), confirming the preferential alignment of precipitate elongation with the external magnetic field.
As the normal interface velocity is proportional to the local driving force, the anisotropy in interfacial fields directly translates into an anisotropic interfacial velocity, thereby enforcing the elongated morphology.
When lattice misfit and elastic coupling are included, the morphology evolves from a smooth ellipsoid to a faceted, elongated cuboid
(Figs.~\ref{fig:evol}(f), (g)-(i)), highlighting the interplay between magnetic and elastic energy contributions. 
Therefore, these distinct morphologies provide compelling evidence of the strong coupled magneto-mechanical interactions dominating the microstructural evolution. 

The influence of magnetic and elastic fields on the precipitate aspect ratio is summarized in Fig.~\ref{fig:aspect}).
We find a sharp increase in the 
aspect ratio from 1 at 0 T to different higher values at 1 T depending on the composition. However, beyond this initial jump from 0 T to 1 T, further increase in either magnetic (Figs.~\ref{fig:aspect}(a)) or elastic field (Figs.~\ref{fig:aspect}(b)-(d)) results in only marginal changes in aspect ratios.
These small changes in the aspect ratios are likely due to the subtle 
adjustments in the shape of the precipitate to accommodate more 
pronounced faceting that occurs with misfit strain shown in 
Fig.~\ref{fig:evol}.

Since we have explicitly considered the anisotropic interfacial field variations within the demagnetization term, the elongation behavior is primarily governed by the demagnetization forces acting at the $\gamma/\alpha$ interface. 
Although the external magnetic field is applied uniformly to the 
entire system, its influence on elongation occurs indirectly through 
modulation of the local interfacial field, rather than through a 
direct shape-selective effect. Since the effective demagnetization 
contribution remains approximately constant across all simulations, 
no substantial field-induced change in aspect ratio is observed.
Nevertheless, an increase in carbon concentration enhances the overall chemical driving force, promoting more isotropic 
growth characteristics and thereby reducing the aspect ratio. This compositional dependence highlights the competition 
between magnetic anisotropy and chemical supersaturation in determining precipitate morphology, underscoring the delicate 
balance between magnetostatic, elastic, and chemical driving forces in field-assisted phase transformations.

As previously discussed, the morphological evolution and resulting characteristics are governed by the coupled chemical, magnetic, and elastic interactions; now, we focus on their influence on the growth behavior (Fig.~\ref{fig:grwrate}).
An increase in the carbon concentration increases the growth rate. 
Increasing the applied field from 0 T to 1 T marginally reduces the growth rate, whereas increasing the magnetic field strength above 1 T significantly reduces it, as shown in  Fig.~\ref{slope1}. 

This arises because the magnetic field introduces anisotropy in the driving force, resulting in growth along the field direction, and it also decreases the magnitude of the effective driving force. Thus, the imposed field acts to constrain isotropic growth, slowing the overall transformation kinetics. When going from 0 to 1 T, the anisotropy dominates, changing the aspect ratio (as shown in Fig.~\ref{fig:aspect}) but not changing the growth rate, and when going from 1 T to higher fields, the decrease in driving force dominates, lowering the growth rate but not having much effect on the aspect ratio. Additionally, the field modifies the local thermodynamic equilibrium at the $\gamma/\alpha$  interface, increasing the equilibrium carbon solubility 
in ferrite~\cite{zhang2005thermodynamic,nakamichi2005diffusion}.
This change reduces the carbon concentration gradient or flux across the interface in the supersaturated ferrite. However, it also increases the transformation of ferrite to austenite per arriving C atoms as the concentration difference is reduced. For the combination of both effects, the reduced diffusion flux usually dominates as long as the equilibrium concentration of austenite is higher than the supersaturated C concentration of ferrite. This results in slower diffusion-limited growth of austenite precipitate.
Furthermore, the applied field shifts the Fe-C phase boundaries to higher transformation temperatures, influencing the 
thermodynamics as well as kinetics of transformation~\cite{zeng2022gibbs}.

The enhancement of growth rate due to increased carbon concentration can be attributed to the increase in the chemical free energy 
difference between matrix and precipitate phases, thereby enhancing the thermodynamic driving force for interface migration. In 
contrast, the application of a magnetic field introduces an additional magnetic free-energy contribution that stabilizes the 
ferromagnetic parent phase and reduces the mobility of the transformation front, effectively slowing the growth. The presence of 
elastic misfit strain further raises the energetic barrier for particle growth by penalizing lattice distortions, amplifying the 
suppressive effect of the field. Overall, these results highlight a three-way competition for growth rate as with the precipitate morphology: carbon concentration enhances growth by 
boosting chemical driving forces, while magnetic field and elastic strain act as counter forces that suppress kinetics, thereby establishing the conditions under which growth is either accelerated or hindered.

In the multi-particle regime, additional mechanisms such as coarsening and coalescence strongly influence both the growth kinetics and the spatial arrangement of precipitates (Fig.~\ref{fig:combr2} and Fig.~\ref{fig:combr3}). 
We can understand this behavior by considering the collective interactions between the precipitates and the magnetic field. As already discussed previously, a $\gamma$‐phase nucleus embedded in the ferromagnetic $\alpha$ matrix behaves as a ``magnetic hole", creating a local magnetic inhomogeneity analogous to the Lorentz‐cavity model in magnetic crystals~\cite{shimotomai2000aligned}. A collection of such holes thus constitutes a many‐body system interacting via dipolar forces similar to nonmagnetic polystyrene particles in a magnetized fluid \cite{skjeltorp1985ordering}. In steel, however, these dipolar interactions occur within a solid matrix and evolve through atomic diffusion~\cite{shimotomai2000aligned}. We can rationalize this behavior using the dipole-dipole interaction in many‐body systems \cite{shimotomai2000aligned,shimotomai2003formation}. 

The dipolar interactions between two spherical particles with magnetic moments $\mathbf{m_1}$, $\mathbf{m_2}$ separated by distance $\mathbf{r}$ can be written as~\cite{cullity2011introduction}:
\begin{equation}
\begin{split}
 U  =   \frac{\mu_0}{4\pi r^3}(\mathbf{m_1} \cdot \mathbf{m_2} - 3(\mathbf{m_1} \cdot \mathbf{\hat{r}})(\mathbf{m_2}\cdot \mathbf{\hat{r}})), 
 \end{split}
 \label{dipole}
\end{equation}
where $\mathbf{\hat{r}}=\mathbf{r}/|r|$ and $\mu_0$ is the permeability of free space.
If both dipoles have the same magnitude 
$m$ and are oriented so that each makes an angle $\theta$ with the line joining them, then Eq.~\ref{dipole} reduces to \cite{skjeltorp1985ordering}:
\begin{equation}
\begin{split}
 U  =   \frac{\mu_0}{4\pi r^3}m^2(1 - 3\cos^2\theta).
 \end{split}
 \label{dipole2}
\end{equation}
According to Eq.~\ref{dipole2} particles aligned parallel to the field ($\theta = 0\degree$) are attracted to each other since $1-3cos^2\theta < 0$, 
promoting directional coalescence. Whereas, those aligned perpendicular 
to the field ($\theta = 90\degree$) are repelled by each other since $1-3\cos^2\theta > 0$, leading to normal coarsening.  
This behavior is consistent with our simulation results.
While the particles grow and align along the direction of the applied magnetic field, on the other hand, the number of particles decreases due to the combined influence of coarsening and coalescence.

\begin{figure}[tbp]
\centering
  \includegraphics[width=1.0
  \linewidth]{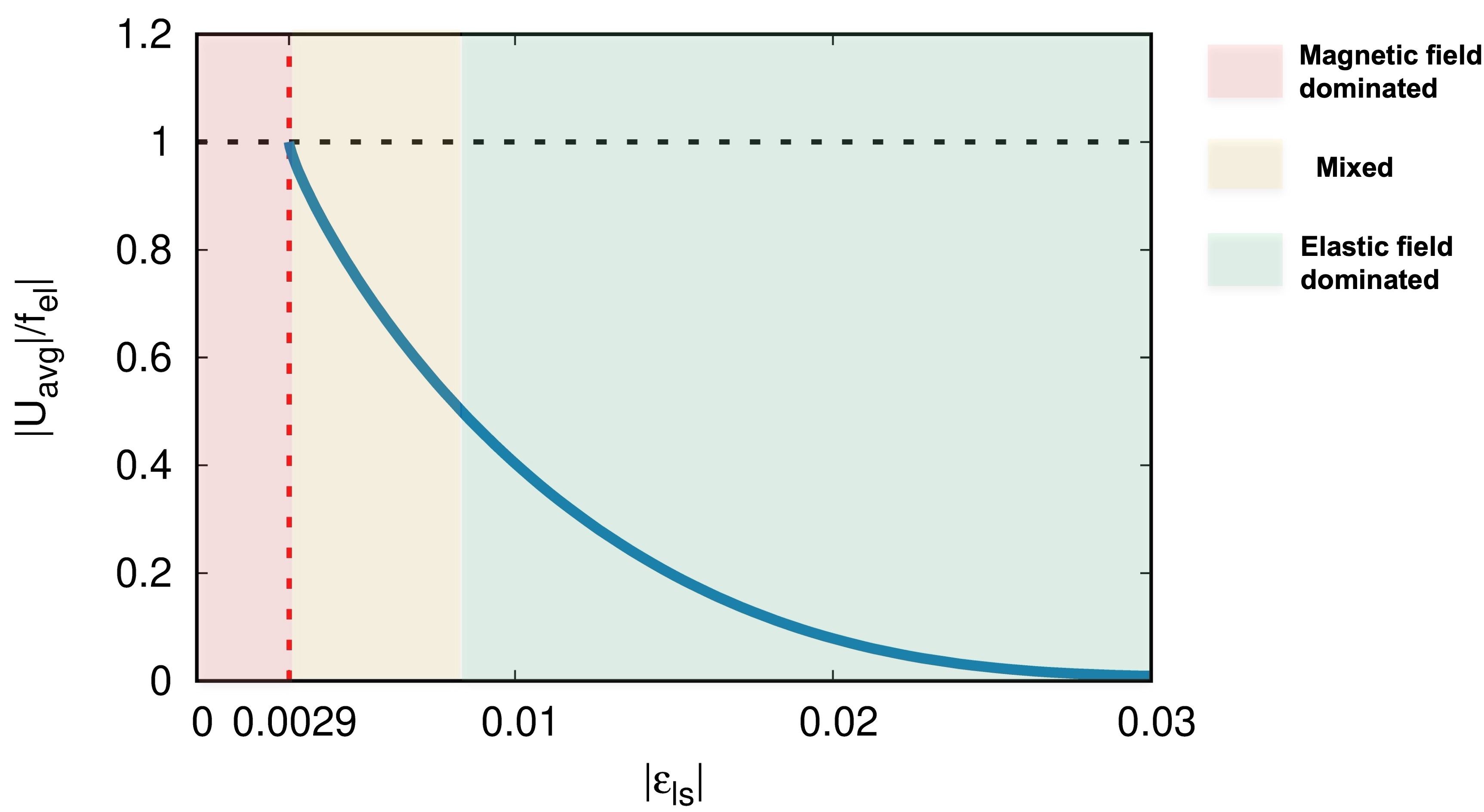}
  \caption{Ratio between the approximate average dipole-dipole and elastic interaction energy densities. This provides a general understanding of the various competing energy regimes.}
\label{fig:dipole_el}
\end{figure}

To develop a general understanding of the relative magnitudes of the competing energies, we compare the approximate dipolar interaction (Eq.~\eqref{dipole2}) with the elastic $f_{el} \approx 0.5G\epsilon_{ls}^2$ contribution. As the dipolar interaction has two components depending on the position of the particles, we consider an average dipolar energy $U_{avg} \approx 0.5(|U_{\theta = 0^\circ}| + U_{\theta=90^\circ})$ to compare with the elastic counterpart. 
Figure~\ref{fig:dipole_el} illustrates the ratio between the average dipole-dipole and elastic energy densities as a function of increasing misfit strain. 

The ratio between the interactions starts at 1, which indicates a crossover between the two competing energy densities occurring at approximately $\epsilon_{ls} \approx 0.0029$, marked by a red dashed line. This crossover signifies the strain at which both 
interactions become comparable in magnitude; below this threshold, magnetic interactions predominantly govern the morphology (red region).  As $\epsilon_{ls}$ increases, the ratio
decreases, indicating increasing dominance of elastic effects (green region). Moreover, we expect a transitional or “mixed” regime (yellow region) to occur within $0.003 < \epsilon_{ls} < 0.008$ (ratio 0.9-0.5), where both interactions compete, jointly influencing the resulting microstructure.
Hence, the overall morphological evolution in this system is governed by a complex interplay of chemical, magnetic, and elastic interactions, highlighting the importance of their coupled effects in determining the final microstructure. 

In addition to these energetic interactions, we observe that a magnetic field-dependent diffusivity further suppresses transformation kinetics (Fig.~\ref{magdiff}). Moreover, an anisotropic diffusivity correction was also examined and found to be minor (less than 2\%) under the present conditions, indicating that diffusion anisotropy plays a negligible role compared with the dominant chemo-magneto-mechanical interactions. Although, we do not include this effect in our present phase field approach, in systems with extremely short diffusion distances (e.g. nanoscale precipitates \cite{wang2024anisotropic}), very steep concentration gradients, strong elastic or other field coupling, or over prolonged aging times, even slight mobility anisotropy can accumulate and lead to measurable shifts in particle aspect ratio or alignment \cite{roy2014growth}.

While this work has provided useful insights into the precipitation in Fe-C alloys subjected to external fields using our novel phase field model, additional enhancements to the model are needed to better predict the behavior of real alloys. First, in this work we introduced stable nuclei at the start of the simulations. In future work, we will incorporate discrete nucleation into the phase field model~\cite{heo2014phase} and investigate the impact of the applied field on classical nucleation theory. In addition, in this work we only considered the growth of $\gamma$ precipitates in an $\alpha$ matrix. In the future we will also consider the nucleation and growth of $\alpha$ precipitates in a $\gamma$ matrix. Finally, in our current model we only considered single crystal $\alpha$ phase and a single $\gamma$ phase orientation. In the future, we will model polycrystalline behavior and incorporate magnetocrystalline anisotropy.

\section{Conclusion} \label{sec:conclusions}
We have established a quantitative process-structure connection for Fe-C alloys subjected to external magnetic fields by coupling CALPHAD‑based chemistry, demagnetization‑field magnetostatics, and microelasticity in a phase‑field framework. The model reproduces classic experimental morphologies (e.g., 8 T alignment), supporting its predictive value for thermomagnetic processing. The external field imposes a directional magnetic driving force that elongates $\gamma$ precipitates along the field, where the elongation originates primarily from the internal demagnetization field concentrated at the $\gamma/\alpha$ interface
while elastic coherency creates faceted, elongated cuboid or brick‑like morphologies under combined magneto‑elastic coupling.  
Growth kinetics increase with carbon content but decrease with magnetic field and misfit strain due to reduced thermodynamic driving force and coherency penalties. Multi‑particle interactions follow dipolar physics: field‑parallel neighbors attract and coalesce directionally, while field‑perpendicular neighbors repel and coarsen conventionally.
There exist a crossover between the magneto-elastic interactions below which the magnetic interaction dominates and beyond that the elastic interaction governs the morphological characteristics.
Incorporating experimentally inferred field‑dependent diffusivity slows kinetics; a magnetostriction‑induced diffusivity anisotropy is estimated to be minor at the conditions considered. Extensions to polycrystalline geometries, explicit nucleation statistics, and magnetocrystalline anisotropy will further enable microstructure‑by‑design in field‑assisted heat treatments.

\section{Acknowledgments}
The authors acknowledge the U.S. Department of Energy’s Office of Energy Efficiency and Renewable Energy (EERE) under the Advanced Manufacturing Office award number DE-EE0009131 for sponsoring the project. We also acknowledge the `High Performance Computation Facility-HiPerGator' at University of Florida for providing the computational resources to perform the simulations. Finally, we are grateful for the leadership of Dr. Michele Manuel on the project. The author (SB) also acknowledges Dr. Vishal Yadav for carefully going through the manuscript and providing valuable feedback.

This report was prepared as an account of work sponsored by an agency of the United States Government.
Neither the United States Government nor any agency thereof, nor any of its employees, makes any warranty,
express or implied, or assumes any legal liability or responsibility for the accuracy, completeness, or
usefulness of any information, apparatus, product, or process disclosed, or represents that its use would not infringe
privately owned rights. Reference herein to any specific commercial product, process, or service by trade name, trademark, manufacturer, or otherwise does not necessarily constitute or imply its
endorsement, recommendation, or
favoring by the
United States Government
or any agency thereof.
The views and opinions of the authors expressed herein do not necessarily state
or reflect those of the United States
Government
or any agency thereof.

\section*{Author Contributions}
\textbf{Soumya Bandyopadhyay:} Conceptualization, Visualization, Methodology, Software, Investigation, Formal analysis, Validation, Data curation, Writing-Original Draft. \textbf{Sourav Chatterjee:} Methodology, Software, Formal analysis. \textbf{Dallas R. Trinkle:} Project administration, Writing - review $\&$ editing, Funding acquisition.
 \textbf{Richard G. Hennig:} Project administration, Writing - review $\&$ editing, Funding acquisition.
  \textbf{Victoria Miller:} Project administration, Writing - review $\&$ editing, Funding acquisition.
  \textbf{Michael S. Kesler:} Project administration, Writing - review $\&$ editing, Funding acquisition.
\textbf{Michael R. Tonks:} Supervision, Project administration, Resources, Writing - review $\&$ editing, Funding acquisition.

\section*{Declaration of Competing Interest}
The authors declare that they have no known competing financial interests or personal relationships that could have appeared to influence the work reported in this article.

\appendix
\section{Determination of the Effective Anisotropy Scaling Factor $\zeta_{mag}$}\label{apnA}

The magnitude of the magnetic driving force in ferromagnetic-paramagnetic systems is largely governed by the magnetostatic or shape-anisotropy energy associated with demagnetizing fields. For a uniformly magnetized ellipsoid, the analytical demagnetization energy density along a principal axis $i$ can be expressed as~\cite{osborn1945demagnetizing,shimotomai2003formation}:
\begin{equation}
    f_{d}^{ana} =  \frac{1}{2}\mu_0 N_{i}M_s^2,
    \label{analyt}
\end{equation}
with $N_x + N_y + N_z = 1$. Here $N_{i}$ are the demagnetization factors determined by the aspect ratio of the inclusion, $M_s=\SI{1.75e6}{\ampere\per\meter}$ is the saturation magnetization, and $\mu_{0} =4\pi\times 10^{-7}$ Hm$^{-1}$ is the magnetic permeability of free space. 

The magnetostatic (shape) anisotropy energy arises from the difference in demagnetizing fields when the magnetization is oriented along the geometrical easy and hard directions of an inclusion, and is expressed as~\cite{osborn1945demagnetizing}:
\begin{equation}
    \Delta f_{d}^{ana} =  \frac{1}{2}\mu_0 \Delta N M_s^2,
    \label{analy}
\end{equation}
where $ \Delta N = N_\perp - N_\parallel$, and from the analytical expressions for prolate spheroids determined by Osborn~\cite{osborn1945demagnetizing}, inclusions with typical aspect ratios of 3:1 - 5:1 yield $ \Delta N \approx 0.33 - 0.43$, which corresponds to the theoretical shape anisotropic energy of $\approx 4 \times 10^{-3} eV/nm^3$ - $5 \times 10^{-3} eV/nm^3$.

These values represent the analytical energy scale associated with 
magnetostatic anisotropy in elongated inclusion in $\gamma/\alpha$ system. However, when magnetization is represented as a 
scalar quantity, 
as in the present formulation, the vectorial nature of 
$\mathbf{m}$ and $\mathbf{h}_d$ and the associated 
anisotropic field variations at curved 
or interfacial regions are not explicitly resolved. Consequently, the 
computed stray-field driving forces are typically an order of 
magnitude lower than these analytical estimates, as the model 
captures only the isotropic component of the demagnetization energy. 
Similar limitations of scalar-magnetization approximations have been 
discussed in prior micromagnetic analyses~\cite{hubert1998magnetic, 
skomski2008simple}.

The unscaled value obtained from the simulations yields a maximum magnetostatic energy density $f_d \approx 2 \times 10^{-4} eV/nm^3$, which is roughly an order of magnitude smaller than the analytical value above. Thus, to restore the correct energetic scale predicted by classical demagnetization theory and to account for the missing anisotropic contributions, the demagnetization term in the free-energy functional was multiplied by an effective anisotropy scaling factor, $\zeta_{mag}$, defined as:
\begin{equation}
\zeta_{mag} \approx \frac{\Delta f_{d}^{ana}}{\Delta f_{d}}
\end{equation}
Since the simulated magnetization is temperature dependent ($M = M_sm(T)$) the corresponding analytical scale at the simulation temperature reduces to $  \Delta f_{d}^{ana} \approx 5.8 \times 10^{-4} eV/nm^3$ - $7.2 \times 10^{-4} eV/nm^3$, yielding an effective ratio $\zeta_{mag} \approx 3 \text{-} 4$. In contrast, when compared with the classical low-temperature reference, the ratio increases to $\zeta_{mag} \approx 20 \text{-} 25$.

Although the temperature adjusted ratio yields a smaller value  $\zeta_{mag} \approx 3 \text{-} 4$, adopting this undercompensates for the attenuation of magnetic driving forces inherent to the scalar magnetization formulation, which already suppresses interfacial vector coupling and shape anisotropy effects.
Therefore, to account for the unresolved 
vectorial and interfacial anisotropy,
we adopted a single representative value
of $\zeta_{mag} = 10$  throughout this work.
This provides a balanced correction lying between the temperature-reduced and low-temperature limits, ensuring that the magnetostatic energy scale remains physically consistent with classical demagnetization behavior.
This choice restores the physically realistic magnetostatic energy scale needed to correlate and comprehend the experimentally observed elongation behavior~\cite{osborn1945demagnetizing,kittel2018introduction,cullity2011introduction}.

\bibliographystyle{unsrt}
\bibliography{scopus}
\clearpage
\end{document}